\documentclass[aps,prd,superscriptaddress,preprintnumbers,twocolumn]{revtex4}

\usepackage{graphicx,amsfonts,amsmath,amssymb,amstext}
\usepackage{float,wrapfig}
\usepackage{subfigure, psfrag}
\usepackage{dsfont}
\usepackage{color}
\usepackage{bm}

\newcommand{\be}{\begin{equation}}
\newcommand{\ee}{\end{equation}}
\newcommand{\ba}{\begin{eqnarray}}
\newcommand{\ea}{\end{eqnarray}}
\newcommand{\sign}[1]{\,\mbox{sgn}\left({#1}\right)}

\definecolor{purple}{rgb}{0.8,0,0.6}
\definecolor{darkgreen}{rgb}{0.00,0.6,0.00}

\begin{document}

\title{Spectroscopic and optical response of odd-frequency superconductors}
\date{August 22, 2019}

\author{P.~O.~Sukhachov}
\email{pavlo.sukhachov@su.se}
\affiliation{Nordita, KTH Royal Institute of Technology and Stockholm University, Roslagstullsbacken 23, SE-106 91 Stockholm, Sweden}

\author{A.~V.~Balatsky}
\affiliation{Nordita, KTH Royal Institute of Technology and Stockholm University, Roslagstullsbacken 23, SE-106 91 Stockholm, Sweden}

\affiliation{Department of Physics, University of Connecticut, Storrs, CT 06269, USA}

\begin{abstract}
The optical response of superconductors with odd-frequency Berezinskii pairing is studied. By using a simple model with a parabolic dispersion law and a non-magnetic disorder, the spectral function, the electron density of states, and the optical conductivity are calculated for a few gap ansatzes. The spectral function and the electron density of states clearly reveal the gap for the Berezinskii pairing for sufficiently strong frequency dependence of the order parameters. It is found that, similarly to the conventional BCS pairing, the odd-frequency gaps induce peaks in the real part of the conductivity, which, however, are sharper than in the BCS case. The magnitude and position of these peaks are determined by the frequency profile of the gap. The imaginary part of the optical conductivity for the Berezinskii pairing demonstrates sharp cusps that are absent in the case of the BCS superconductors. The corresponding results suggest that the Berezinskii pairing might allow for the optical transparency windows related to the onsets of the attenuation peaks in the real part of the conductivity. Thus, the study of the optical response not only provides an alternative way to probe the odd-frequency gaps but can reveal also additional features of the dynamic superconducting pairing.
\end{abstract}

\maketitle

\section{Introduction}
\label{sec:Introduction}

Odd-frequency (OF) superconductivity was suggested by Berezinskii in the 1970s as a possible order parameter for superfluid He$^3$~\cite{Berezinskii:1974}. He pointed out that the pairing states could have specific symmetry properties, which later were shown to be classified via the so-called $SP^{*}OT^{*}$ rule, for a review see Ref.~\cite{Linder-Balatsky:rev-2017}.
This rule clarifies the symmetry of the superconducting gap under the operation of the spin-permutation $S$, coordinate-inversion $P^{*}$, orbital-interchange $O$, and time-permutation $T^{*}$. Symbolically, the rule can be written as
\begin{equation}
\label{intro-SPOT}
SP^{*}OT^{*} = -1
\end{equation}
and provides a transparent way to classify the superconducting gaps with respect to their symmetry properties without knowing microscopic details of the pairing. The crucial ingredient for the OF pairing in Eq.~(\ref{intro-SPOT}) is the time-permutation $T^{*}$ symmetry. This latter was used by Berezinskii to suggest a spin-triplet ($S=+1$) s-wave pairing ($P^{*}=+1$) channel in He$^3$ (where the orbital index is $O=+1$). Although the odd-frequency pairing channel in He$^3$  was not confirmed, the concept of the nonlocal in time pairing proved to be a fruitful theoretical idea.

In the 1990's the concept of the OF or, equivalently, Berezinskii pairing was rekindled in the context of superconductivity by Balatsky and Abrahams~\cite{Balatsky-Abrahams:1992,Abrahams-Allen:1993}. There are numerous examples that the OF superconducting states might naturally appear in various condensed matter systems~\cite{Linder-Balatsky:rev-2017}. The often mentioned examples include superconducting heterostructures, where the Berezinskii OF component can be induced as a result of scattering on either normal or ferromagnetic material brought in the contact with the superconductor~\cite{Tanaka-Kashiwaya:2005,Asano-Golubov:2007,Tanaka-Golubov:2007a,Tanaka-Golubov:2007b,Eschrig-Schon:2007}. Similar effect was predicted to appear also on the boundary with the unconventional superconductors~\cite{Tanaka-Golubov:2007c,Asano-Tanaka:2011,Matsumoto-Kusunose:2013,Lu-Tanaka:2016}. The OF superconducting state might appear near the Abrikosov vortex core in the type-II superconductors~\cite{Yokoyama-Golubov:2008,Tanuma-Golubov:2009,Yokoyama-Tanaka:2010,Daino-Tanaka:2012,Bjornson-Black-Schaffer:2015,Tanaka-Onari:2016}. It is natural to expect the Berezinskii pairing in the systems that experience a time-dependent drive~\cite{Triola-Balatsky:2016,Triola-Balatsky:2017}.

As is evident from the $SP^{*}OT^{*}$ rule (\ref{intro-SPOT}), another potentially viable possibility to achieve the Berezinskii-type pairing, is to consider various multiband systems~\cite{Black-Schaffer-Balatsky:2013,Komendova-Black-Schaffer:2015,Asano-Sasaki:2015,Triola-Balatsky:2017,Triola-Black-Schaffer:rev-2019}. Another platform where OF pairing might naturally appear in Dirac and Weyl semimetals~\cite{Sukhachov-Balatsky:2019}, where chirality of Weyl or Dirac nodes plays the role of the band index.

Recently, topological superconductors appeared as a new platform to realize unconventional types of the pairings. (For reviews on topological superconductivity, see Refs.~\cite{Bernevig,Schnyder-rev:2015,Sato:2016evq}.) It was demonstrated that the zero-energy quasiparticle states in topological superconductors, known as the Majorana states~\cite{Alicea:2012,Beenakker:2013,Stanescu:2013}, provide a natural way to achieve a mixed pairing state, where both the even- and odd-frequency components are simultaneously present~\cite{Huang-Balatsky:2015}.

Until now, there is no definite experimental evidence for the OF superconductivity in the ground state. Spectroscopic signatures of OF Cooper pairs were seen in the density of states (DOS)~\cite{Di-Bernardo-Robinson:2015} as well as the possible paramagnetic Meissner response ~\cite{Di-Bernardo-Robinson:2015a}. It is worth noting, however, that the latter is strongly debated (see e.g., Refs.~\cite{Solenov-Mozyrsky:2009,Kusunose-Miyake:2011}) and should be considered with caution.

The notion of OF pairing is a relatively new concept and complete understanding of this pairing and its properties is still developing. To facilitate better understanding of OF state and provide a better guidance of Berezinskii pairing, we calculate both the local DOS and spectroscopic features like optical conductivity in clean and disordered Berezinskii superconductors for a few choices of OF gaps. To the best of our knowledge, optical conductivity of the OF pairing state was not investigated before.

Electronic transport and optical response are the often used tools to detect transitions and onset of superconducting state in conventional superconductors~\cite{Schrieffer:book,Abrikosov-Dzyaloshinski:book}. The real part of the optical conductivity describes the attenuation of the electromagnetic waves and the imaginary one quantifies the dielectric properties.
It is clear that the corresponding study is important not only from a theoretical point of view but could be useful for the experimental identification of OF Cooper pairs.

By using a simple model of OF superconductors with a parabolic dispersion law and a non-magnetic disorder, which nevertheless captures the key features of the Berezinskii pairing, we calculate the optical conductivity in the standard Kubo linear response approach. It is found that the OF pairing allows for the peaks in the real part of the optical conductivity, whose form and position depend on the specific frequency dependence of the gap. For the frequencies that correspond to these peaks, the sharp cusps appear in the imaginary part, as is indeed expected from the Kramers--Kronig relations. The form of these cusps is, however, nontrivial for the OF pairing. In particular, the imaginary part of the optical conductivity can become negative allowing for the optical transparency windows. We believe that these predictions will stimulate the experimental search and identification of the Berezinskii pairing via optical probes.
In addition to the optical conductivity, the spectral function and the electron DOS are also obtained. While they clearly demonstrate the generation of the spectral gap and coherence peaks for the sufficiently strong frequency dependence of the gaps, the corresponding results lack universal signatures that can be used to unambiguously distinguish odd- and even-frequency pairing channels.

The paper is organized as follows. We introduce the model of an OF superconductor and provide the key definitions in Sec.~\ref{sec:Model}. A few ansatzes are assumed for the OF gaps. Sec.~\ref{sec:DOS} is devoted to the spectral function and the electron DOS. The optical response is calculated and the corresponding results for the real and imaginary parts of the conductivity are given in Sec.~\ref{sec:OC-calc}. Our results are discussed and summarized in Sec.~\ref{sec:Summary}. Technical details including the optical sum rule are presented in Appendix~\ref{sec:App-Sum-rules}.
Throughout this paper, we set $\hbar=k_B=1$.

\section{Model}
\label{sec:Model}

In this section, we present the minimal model of the OF pairing as well as provide the key definitions that will be used in the analysis of the spectral function, the DOS, and the optical conductivity below.

\subsection{Green's functions in the Nambu space}
\label{sec:Model-Green}

Since the OF pairing is a dynamic quantum order that is essentially nonlocal in time, the corresponding mean-field Hamiltonian cannot be formulated~\cite{Solenov-Mozyrsky:2009,Kusunose-Miyake:2011}. By using the effective action approach for the OF superconductors elaborated in Refs.~\cite{Solenov-Mozyrsky:2009,Kusunose-Miyake:2011} the retarded (advanced) Green function in the Nambu space reads
\begin{eqnarray}
\label{OC-SF-G-BdG-def}
\hat{S}^{R/A}(\omega,\mathbf{p}) &=& \Big[Z(\omega, \mathbf{p})(\omega \pm i0) \mathds{1}_2 -\hat{H}_{\rm N} \nonumber\\
&-&Z(\omega, \mathbf{p}) \hat{\Delta}(\omega,\mathbf{p})\Big]^{-1},
\end{eqnarray}
where $\mathds{1}_2$ is the $2\times 2$ matrix in the Nambu space and
\begin{eqnarray}
\label{OC-H-def}
\hat{H}_{\rm N} = \left(
                \begin{array}{cc}
                  \xi & 0 \\
                  0 & -\xi \\
                \end{array}
              \right)
\end{eqnarray}
is the free Hamiltonian in the Nambu space. In addition, $\omega$ is the frequency, $\xi=p^2/(2m)-\mu$ is the quasiparticle energy with respect to the Fermi level, $\mathbf{p}$ is the momentum, $\mu$ is the electric chemical potential, and $m$ is the mass of a charge carrier. The gap matrix $\hat{\Delta}(\omega,\mathbf{p})$ is defined as
\begin{eqnarray}
\label{OC-Delta-def}
\hat{\Delta}(\omega,\mathbf{p}) = \left(
                \begin{array}{cc}
                  0 & \Delta(\omega,\mathbf{p}) \\
                  \Delta^{\dag}(\omega,\mathbf{p}) & 0 \\
                \end{array}
              \right).
\end{eqnarray}
The physical meaning of the coefficient $Z(\omega, \mathbf{p})$ will be clarified in Sec.~\ref{sec:DOS}. Here we note that it might appear in the Green function due to the wave function renormalization when the generation of the OF gap is considered self-consistently. In this study, however, it will be treated phenomenologically.

By inverting Eq.~(\ref{OC-SF-G-BdG-def}), the explicit form of the Green function in the Nambu space reads
\begin{widetext}
\begin{eqnarray}
\label{OC-SF-G-BdG-def-1}
\hat{S}^{R/A}(\omega,\mathbf{p}) = \frac{1}{Z(\omega, \mathbf{p})} \frac{\omega\mathds{1}_2 +\xi\tau_z/Z(\omega, \mathbf{p})+ \Delta(\omega,\mathbf{p}) \tau_{+}+ \Delta^{\dag}(\omega,\mathbf{p})\tau_{-}}{\omega^2-\xi^2/Z^2(\omega, \mathbf{p}) -|\Delta(\omega,\mathbf{p})|^2 \pm i0\sign{\omega}},
\end{eqnarray}
\end{widetext}
where $\bm{\tau}$ are the Pauli matrices that act in the Nambu space and $\tau_{\pm}=\tau_x \pm i\tau_y$. Note that the gap $\Delta(\omega,\mathbf{p})$ is a function of both frequency and momentum and, in general, can describe both even- and odd-frequency pairings of different symmetries (e.g., s- or p-wave pairings).

\subsection{Gap ansatzes}
\label{sec:Model-gaps}

\begin{figure*}[!ht]
\begin{center}
\includegraphics[width=0.45\textwidth]{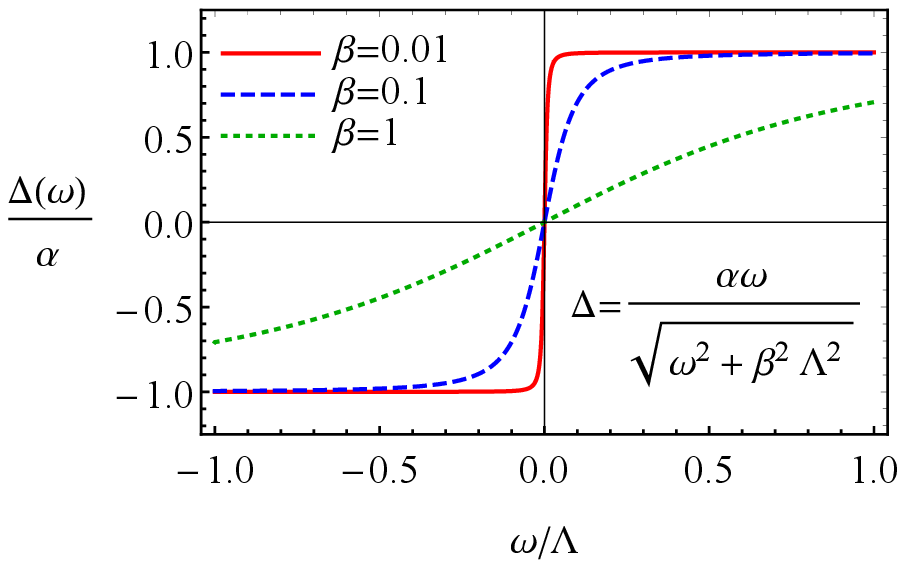}\hfill
\includegraphics[width=0.45\textwidth]{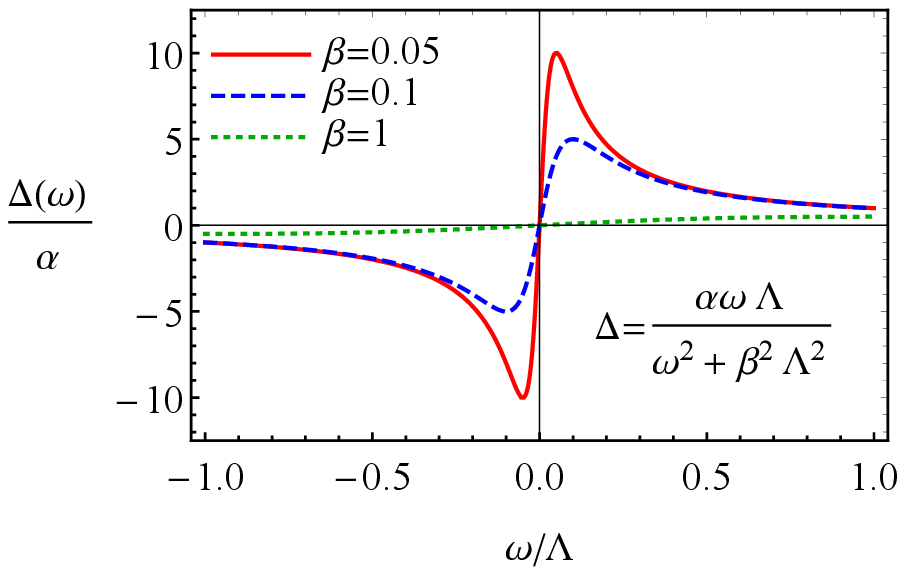}
\end{center}
\hspace{0.075\textwidth}{\small (a)}\hspace{0.525\textwidth}{\small (b)}\\[0pt]
\caption{The dependence of the s-wave OF gaps $\Delta=\alpha \omega/\sqrt{\omega^2 +\beta^2 \Lambda^2}$ (panel (a)) and $\Delta=\alpha \omega \Lambda/\left(\omega^2 +\beta^2 \Lambda^2\right)$ (panel (b)) on frequency $\omega$ for a few values of the control parameter $\beta$. 
}
\label{fig:Disorder-SP-gaps}
\end{figure*}

As the starting point in our analysis of the OF pairing, we use two reasonable anzatses for the s-wave OF gap, i.e.,
\begin{eqnarray}
\label{Model-gaps-Delta-1}
\Delta(\omega,\mathbf{p})&=&\frac{\alpha \omega}{\sqrt{\omega^2 +\beta^2 \Lambda^2}},\\
\label{Model-gaps-Delta-2}
\Delta(\omega,\mathbf{p})&=&\frac{\alpha \omega \Lambda}{\omega^2 +\beta^2 \Lambda^2}.
\end{eqnarray}
These gaps at a few different values of the control parameter $\beta$ are shown in Figs.~\ref{fig:Disorder-SP-gaps}(a) and \ref{fig:Disorder-SP-gaps}(b), respectively. It is clear that the gap given in Eq.~(\ref{Model-gaps-Delta-1}) has a step-like dependence on the frequency for small values of $\beta$. Indeed, it reduces to one of the simplest OF gaps at $\beta\to0$, i.e., $\Delta(\omega)\to\alpha \sign{\omega}$. On the other hand, the gap in Eq.~(\ref{Model-gaps-Delta-2}) has a nonmonotonic dependence on $\omega$, where the peak-like features are clearly seen for small values of $\beta$. It is worth noting that the gaps of a different symmetry can be described by replacing $\alpha$ with a suitable momentum-dependent function. For example, the p-wave pairing can be described by redefining $\alpha\to\alpha \cos{\varphi}$ where $\varphi=\arctan{\left(p_y/p_x\right)}$.

\subsection{Effects of disorder}
\label{sec:Model-disorder}

Disorder plays an important role in transport properties of any material. Since it is ubiquitously present in real samples, it is convenient to include the corresponding effects in our study of the optical response. By following the standard textbook approach (see, e.g., Refs.~\cite{Abrikosov-Dzyaloshinski:book,Schrieffer:book,Levitov:book}, the effects of impurities are taken via the self-energy correction to Green's function as
\begin{eqnarray}
\label{Disorder-SE-G-full-def}
\hat{G}^{-1}(\omega,\mathbf{p}) = \hat{S}^{-1}(\omega,\mathbf{p}) -\hat{\Sigma}(\omega,\mathbf{p}),
\end{eqnarray}
where  $\hat{S}^{-1}(\omega,\mathbf{p})$ is the inverse free Green function, which straightforwardly follows from Eqs.~(\ref{OC-SF-G-BdG-def}) or (\ref{Disorder-SE-G-full-def}). Further, $\hat{G}^{-1}(\omega,\mathbf{p})$ is the inverse full Green function. The self-energy $\hat{\Sigma}(\omega,\mathbf{p})$ in the fist-order (Born) approximation reads as
\begin{equation}
\label{Disorder-SE-Sigma-def}
\hat{\Sigma}(\Omega, \mathbf{q}) = \int d\omega \int \frac{d^np}{(2\pi)^{n+1}} \tau_z \hat{S}(\omega, \mathbf{p}) \hat{D}(\Omega-\omega,\mathbf{q}-\mathbf{p}) \tau_z,
\end{equation}
where $\hat{D}(\Omega-\omega,\mathbf{q}-\mathbf{p})$ is the disorder propagator and $n=2,3$ is the spacial dimension of the superconductor. For simplicity, let us consider the case of a non-magnetic disorder with the following propagator:
\begin{eqnarray}
\label{Disorder-SE-D-def}
\hat{D}(\omega,\mathbf{p}) = 2\pi\delta(\omega) n_{\rm imp}u_0^2.
\end{eqnarray}
Here $u_0$ is the strength of the disorder potential and $n_{\rm imp}$ is the concentration of impurities. It is clear that such disorder does not change the pairing state of the electrons. In addition, we employ the conventional approximation valid in material with a large Fermi surface, i.e.,
\begin{eqnarray}
\label{Disorder-SE-xi-int}
\int \frac{d^n p}{(2\pi)^n} \to \frac{\nu_0}{4\pi} \int_{-\Lambda}^{\Lambda} d\xi \int d\Omega_{\mathbf{p}}.
\end{eqnarray}
Here $\Lambda$ in the energy scale, $\nu_0$ is the density of states at the Fermi level,
\begin{eqnarray}
\label{Disorder-SE-nu-0}
\mbox{2D:} \quad \nu_0 &=& \frac{m}{2\pi},\\
\mbox{3D:} \quad  \nu_0 &=& \frac{m\sqrt{2m\mu}}{2\pi^2},
\end{eqnarray}
and $\int d\Omega_{\mathbf{p}}$ denotes the integration over angles, i.e.,
\begin{eqnarray}
\label{Disorder-SE-Omega-p}
\mbox{2D:} \quad \int d\Omega_{\mathbf{p}} &=& \int_0^{2\pi} d\varphi,\\
\mbox{3D:} \quad \int d\Omega_{\mathbf{p}} &=& \int_0^{\pi} d\theta\,\sin{\theta} \int_0^{2\pi} d\varphi.
\end{eqnarray}

Then, the self-energy in Eq.~(\ref{Disorder-SE-Sigma-def}) can be rewritten as
\begin{eqnarray}
\label{Disorder-SE-Sigma}
\hat{\Sigma}^{\rm A/R} &=& \frac{1}{2\pi \nu_0 \tau} \frac{\nu_0}{4\pi} \int d\Omega_{\mathbf{p}} \int_{-\Lambda}^{\Lambda} d \xi \frac{1}{Z(\omega, \mathbf{p})} \nonumber\\
&\times&  \frac{\omega \mathds{1}_2 +\xi \tau_z/Z(\omega, \mathbf{p}) -\Delta(\omega,\mathbf{p}) \tau_{+} -\Delta^{\dag}(\omega,\mathbf{p}) \tau_{-}}{\omega^2 -\xi^2/Z^2(\omega, \mathbf{p}) - |\Delta(\omega,\mathbf{p})|^2 \mp i0\sign{\omega}}, \nonumber\\
\end{eqnarray}
where we introduced the relaxation time $\tau$ as follows:
\begin{eqnarray}
\label{Disorder-SE-tau-def}
\tau \equiv \frac{1}{2\pi \nu_0 n_{\rm imp} u_0^2}.
\end{eqnarray}
By calculating the integral over $\xi$, we obtain
\begin{eqnarray}
\label{Disorder-SE-Sigma-Re}
\mbox{Re}\, \hat{\Sigma}^{\rm A/R} &=& -\frac{1}{2\tau} \frac{1}{4\pi} \int d\Omega_{\mathbf{p}} \frac{\theta\left(|\Delta(\omega,\mathbf{p})|^2-\omega^2\right)}{\sqrt{|\Delta(\omega,\mathbf{p})|^2-\omega^2}}\nonumber\\
&\times&\left[\omega \mathds{1}_2 -\Delta(\omega,\mathbf{p}) \tau_{+} -\Delta^{\dag}(\omega,\mathbf{p}) \tau_{-}\right]
\end{eqnarray}
and
\begin{eqnarray}
\label{Disorder-SE-Sigma-Im}
\mbox{Im}\, \hat{\Sigma}^{\rm A/R} &=& \pm\frac{\sign{\omega}}{2\tau} \frac{1}{4\pi} \int d\Omega_{\mathbf{p}} \frac{\theta\left(\omega^2-|\Delta(\omega,\mathbf{p})|^2\right)}{\sqrt{\omega^2-|\Delta(\omega,\mathbf{p})|^2}} \nonumber\\
&\times&\left[\omega \mathds{1}_2 -\Delta(\omega,\mathbf{p}) \tau_{+} -\Delta^{\dag}(\omega,\mathbf{p}) \tau_{-}\right].
\end{eqnarray}
Here $\theta(x)$ is the Heaviside step function. For simplicity, we assumed that $\Lambda\to\infty$ and the dependence of $Z(\omega, \mathbf{p})$ and $\Delta(\omega, \mathbf{p})$ on $\xi$ is weak. Since the relaxation time $\tau$ is a free parameter of the model, we believe that such an approximation should be correct at least qualitatively.

Then, by using Eq.~(\ref{Disorder-SE-G-full-def}), we obtain the following Green's function:
\begin{widetext}
\begin{eqnarray}
\label{Disorder-SE-G-full}
\hat{G}^{\rm A/R}(\omega,\mathbf{p}) &=& \left[Z(\omega, \mathbf{p}) \tilde{\omega} \mathds{1}_2 -\xi \tau_z -Z(\omega, \mathbf{p})\tilde{\Delta}(\omega,\mathbf{p}) \tau_{+} -Z(\omega, \mathbf{p})\tilde{\Delta}^{\dag}(\omega,\mathbf{p}) \tau_{-}\right]^{-1} \nonumber\\
&=& \frac{1}{Z(\omega, \mathbf{p})} \frac{\left[\omega +\xi \tau_z/Z(\omega, \mathbf{p}) +\Delta(\omega) \tau_{+} +\Delta^{\dag}(\omega,\mathbf{p}) \tau_{-}\right] \eta_{\rm Re} \mp i \eta_{\rm Im} \left[\omega +\Delta(\omega,\mathbf{p})\tau_{+} +\Delta^{\dag}(\omega) \tau_{-}\right]}{\left[\omega^2 -|\Delta(\omega,\mathbf{p})|^2\right] \left(\eta_{\rm Re}^2 - \eta_{\rm Im}^2\right) -\xi^2/Z^2(\omega, \mathbf{p}) \mp 2i\eta_{\rm Re}\eta_{\rm Im}\left[\omega^2 -|\Delta(\omega,\mathbf{p})|^2\right]}.
\end{eqnarray}
\end{widetext}
where $\tilde{\omega} = \omega \eta$, $\tilde{\Delta}(\omega,\mathbf{p}) = \Delta(\omega,\mathbf{p}) \eta$, and $\tilde{\Delta}^{\dag}(\omega,\mathbf{p}) = \Delta^{\dag}(\omega,\mathbf{p}) \eta$. The coefficient $\eta$ contains real and imaginary part, $\eta=\eta_{\rm Re}\mp i\eta_{\rm Im}$, where
\begin{eqnarray}
\label{Disorder-SE-eta-Re}
\eta_{\rm Re} &=& 1 +\frac{1}{2\tau Z(\omega, \mathbf{p})} \int \frac{d\Omega_{\mathbf{p}}}{4\pi} \frac{\theta\left(|\Delta(\omega,\mathbf{p})|^2-\omega^2\right)}{\sqrt{|\Delta(\omega,\mathbf{p})|^2-\omega^2}},\nonumber\\
\\
\label{Disorder-SE-eta-Im}
\eta_{\rm Im} &=& \frac{\sign{\omega}}{2\tau Z(\omega, \mathbf{p})} \int \frac{d\Omega_{\mathbf{p}}}{4\pi} \frac{\theta\left(\omega^2-|\Delta(\omega,\mathbf{p})|^2\right)}{\sqrt{\omega^2-|\Delta(\omega,\mathbf{p})|^2}}.
\end{eqnarray}
Note that the sign $-$ ($+$) in $\eta$ corresponds to the advanced (retarded) Green function $\hat{G}^{\rm A}(\omega,\mathbf{p})$ ($\hat{G}^{\rm R}(\omega,\mathbf{p})$). In the case of s-wave superconductivity, the gap does not depend on the angles. Therefore, the corresponding integrals can be trivially taken and give $2\pi$ in 2D or $4\pi$ in 3D. The case of p-wave gaps, where $\alpha\to\alpha \cos{\varphi}$, is more complicated and does not allow for simple analytical expressions. Technically, this is related to the unit step functions.

It is worth noting that the Anderson theorem~\cite{Anderson:1959}
\begin{eqnarray}
\label{Disorder-SE-tilde-omega-omega}
\frac{\tilde{\omega}}{\tilde{\Delta}(\omega,\mathbf{p})} = \frac{\omega}{\Delta(\omega,\mathbf{p})}
\end{eqnarray}
holds in the case under consideration. One can verify that the obtained results are valid, at least within the range of used approximations, even if the self-energy is considered self-consistently (i.e., with the full Green function $G(\omega,\mathbf{p})$ instead of $S(\omega,\mathbf{p})$ in Eq.~(\ref{Disorder-SE-Sigma-def})). Indeed, as follows from Eqs.~(\ref{Disorder-SE-Sigma-Re}) and (\ref{Disorder-SE-Sigma-Im}), the self-energy does not change when the terms with $\omega$, $\Delta$, and $\Delta^{\dag}$ are rescaled simultaneously.

\section{Spectral function and density of states}
\label{sec:DOS}

In this section, we discuss possible signatures of the Berezinskii pairing in the spectral function and the electron DOS. Indeed, these quantities are of outmost importance for various spectroscopic probes such as the angle-resolved photoemission and scanning tunneling spectroscopies. Therefore, it is important to clarify whether the even- and odd-frequency parings can be easily distinguished by such direct probes.

\subsection{Spectral function}
\label{sec:DOS-A}

The spectral function is defined as~\cite{Schrieffer:book,Mahan:book,Bruus-book}
\begin{eqnarray}
\label{Disorder-SP-A-def}
\hat{A}(\omega,\mathbf{p}) = \frac{1}{2\pi i} \left[\hat{G}^{\rm A}(\omega,\mathbf{p}) -\hat{G}^{\rm R}(\omega,\mathbf{p})\right].
\end{eqnarray}
In the clean case $\tau\to\infty$ the corresponding expression reads as
\begin{eqnarray}
\label{Disorder-SP-A-clean}
\hat{A}(\omega,\mathbf{p}) &=& \sign{\omega}\Big[Z(\omega, \mathbf{p}) \omega+\xi\tau_z \nonumber\\
&+&Z(\omega, \mathbf{p})\Delta(\omega,\mathbf{p}) \tau_{+}+Z(\omega, \mathbf{p})\Delta^{\dag}(\omega,\mathbf{p})\tau_{-}\Big] \nonumber\\
&\times&\delta\left(Z^2(\omega, \mathbf{p}) \omega^2-\xi^2 -Z^2(\omega, \mathbf{p})|\Delta(\omega,\mathbf{p})|^2\right).\nonumber\\
\end{eqnarray}
The spectral function for dirty superconductors is more complicated, i.e.,
\begin{widetext}
\begin{eqnarray}
\label{Disorder-SP-A}
\hat{A}(\omega,\mathbf{p}) &=& \frac{1}{\pi Z(\omega, \mathbf{p})} \frac{2 \eta_{\rm Re}^2 \eta_{\rm Im} \left[\omega^2-|\Delta(\omega,\mathbf{p})|^2\right]\left[\omega \mathds{1}_2 +\xi \tau_z/Z(\omega, \mathbf{p}) +\Delta(\omega,\mathbf{p}) \tau_{+} +\Delta^{\dag}(\omega,\mathbf{p}) \tau_{-}\right]}{\left\{\left[\omega^2 -|\Delta(\omega,\mathbf{p})|^2\right] \left(\eta_{\rm Re}^2 -\eta_{\rm Im}^2\right) -\xi^2/Z^2(\omega, \mathbf{p})\right\}^2 +\left\{2\left[\omega^2 -|\Delta(\omega,\mathbf{p})|^2\right] \eta_{\rm Re} \eta_{\rm Im}\right\}^2} \nonumber\\
&-& \frac{1}{\pi Z(\omega, \mathbf{p})} \frac{\eta_{\rm Im} \left[\omega \mathds{1}_2 +\Delta(\omega,\mathbf{p}) \tau_{+} +\Delta^{\dag}(\omega,\mathbf{p}) \tau_{-}\right] \left\{\left[\omega^2 -|\Delta(\omega,\mathbf{p})|^2\right] \left(\eta_{\rm Re}^2 - \eta_{\rm Im}^2\right) -\xi^2/Z^2(\omega, \mathbf{p})\right\} }{\left\{\left[\omega^2 -|\Delta(\omega,\mathbf{p})|^2\right] \left(\eta_{\rm Re}^2 -\eta_{\rm Im}^2\right) -\xi^2/Z^2(\omega, \mathbf{p})\right\}^2 +\left\{2\left[\omega^2 -|\Delta(\omega,\mathbf{p})|^2\right] \eta_{\rm Re} \eta_{\rm Im}\right\}^2}.
\end{eqnarray}
\end{widetext}
Note that Green's function can be easily restored via known spectral function as
\begin{eqnarray}
\label{Disorder-SP-G-A}
\hat{G}(\Omega,\mathbf{p}) = \int \frac{d\omega \hat{A}(\omega,\mathbf{p})}{\Omega-\omega}.
\end{eqnarray}

\begin{figure*}[!ht]
\begin{center}
\includegraphics[width=0.45\textwidth]{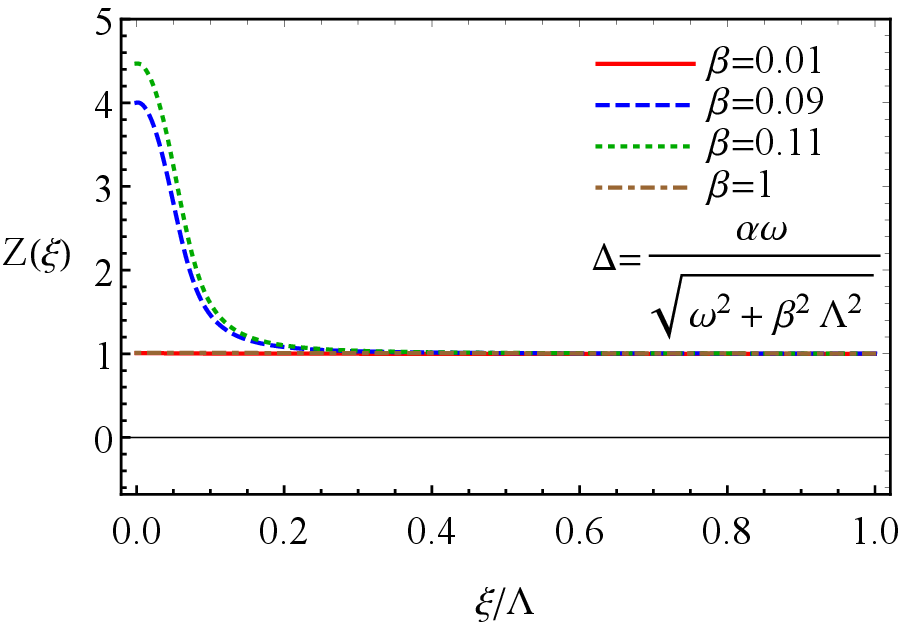}\hfill
\includegraphics[width=0.45\textwidth]{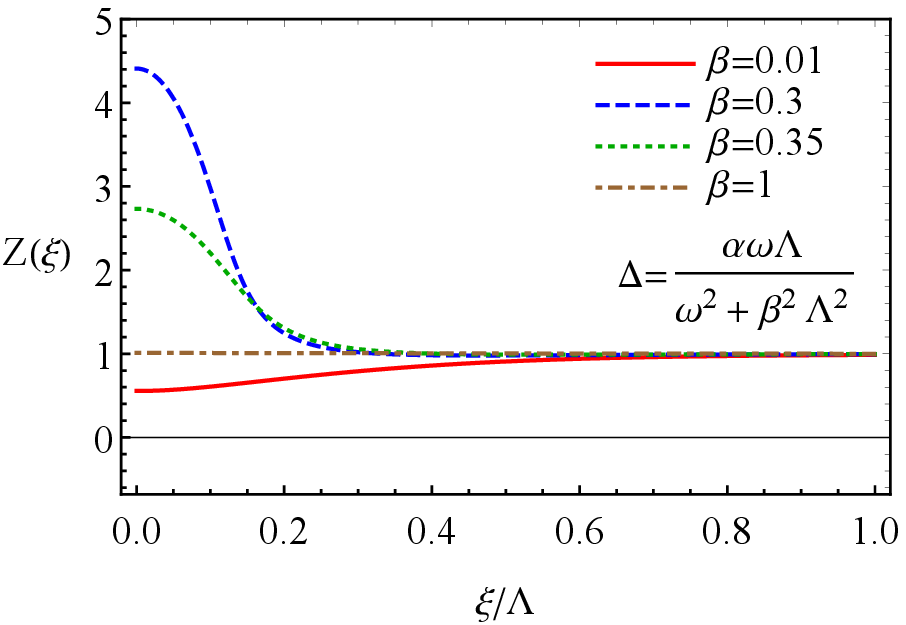}
\end{center}
\hspace{0.075\textwidth}{\small (a)}\hspace{0.525\textwidth}{\small (b)}\\[0pt]
\caption{The coefficient $Z(\omega, \mathbf{p}) \approx Z(\xi)$ as a function of the energy $\xi$ for $\Delta=\alpha \omega/\sqrt{\omega^2 +\beta^2 \Lambda^2}$ (panel (a)) and $\Delta=\alpha \omega \Lambda/\left(\omega^2 +\beta^2 \Lambda^2\right)$ (panel (b)). In both panels we set $\tau\to 10/\Lambda$, $\alpha=0.1\Lambda$, and $T=0$.
}
\label{fig:Z-s-wave-4-2-eta}
\end{figure*}

Before proceeding with the characteristic features of the Berezinskii pairing, let us discuss the role of the coefficient $Z(\omega, \mathbf{p})$. According to Refs.~\cite{Bruus-book,Altland:book,Mahan:book}, the electron part of the spectral function $\hat{A}(\omega,\mathbf{p})$ in the Nambu space integrated over all frequencies, i.e.,
\begin{eqnarray}
\label{OC-sum-rules-IA}
I_{\rm A} = \int_{-\infty}^{\infty}d\omega\, \mbox{tr}\left[\frac{1+\tau_z}{2}\hat{A}(\omega,\mathbf{p})\right],
\end{eqnarray}
should satisfy the following sum rule:
\begin{eqnarray}
\label{OC-sum-rules-AG}
I_{\rm A} = 1.
\end{eqnarray}
It is straightforward to check that neither conventional BCS $\Delta(\omega)=\alpha$ nor simplest odd-frequency gap $\Delta(\omega)=\alpha\sign{\omega}$ breaks the sum rule (\ref{OC-sum-rules-AG}) at $Z(\omega, \mathbf{p})=1$. One can check, however, that the OF gaps $\Delta=\alpha \omega/\sqrt{\omega^2 +\beta^2 \Lambda^2}$ and $\Delta=\alpha \omega \Lambda/\left(\omega^2 +\beta^2 \Lambda^2\right)$ generically break the sum rule if $Z(\omega, \mathbf{p})$ is ignored (see the results in Appendix~\ref{sec:App-Sum-rules}). The deviations might be severe at certain values of $\beta$ where the spectral gap closes. Therefore, in order to mitigate the breakdown, we introduced the additional coefficient $Z(\omega, \mathbf{p})$, which could originate from the self-consistent treatment of the gap similarly to the conventional Eliashberg approach. Since the self-consistent approach is not employed in this study, we determine $Z(\omega, \mathbf{p})$ from the sum rule itself. In order to simplify the solution of the complicated integral equation, it is reasonable to assume that $Z(\omega, \mathbf{p}) \approx Z(\xi)$. Then, one can proceed with the standard iterative methods. The results for the coefficient $Z(\xi)$ in the case of dirty superconductors are shown in Figs.~\ref{fig:Z-s-wave-4-2-eta}(a) and \ref{fig:Z-s-wave-4-2-eta}(b) for $\Delta=\alpha \omega/\sqrt{\omega^2 +\beta^2 \Lambda^2}$ and $\Delta=\alpha \omega \Lambda/\left(\omega^2 +\beta^2 \Lambda^2\right)$, respectively. The same results, albeit in the clean case are presented in Figs.~\ref{fig:Z-s-wave-4-2}(a) and \ref{fig:Z-s-wave-4-2}(b) in Appendix~\ref{sec:App-Sum-rules}. Note that the difference between the clean and dirty cases is quantitative rather than qualitative.

\begin{figure*}[!ht]
\begin{center}
\includegraphics[width=0.45\textwidth]{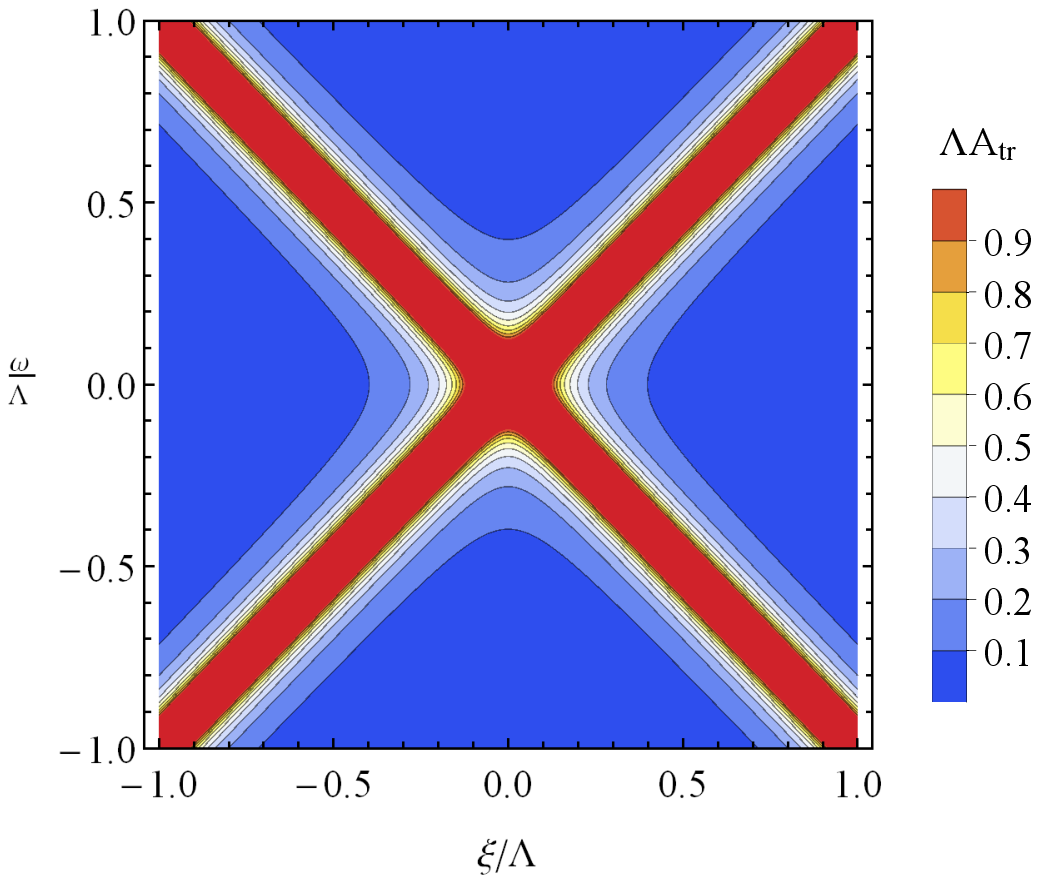}\hfill
\includegraphics[width=0.45\textwidth]{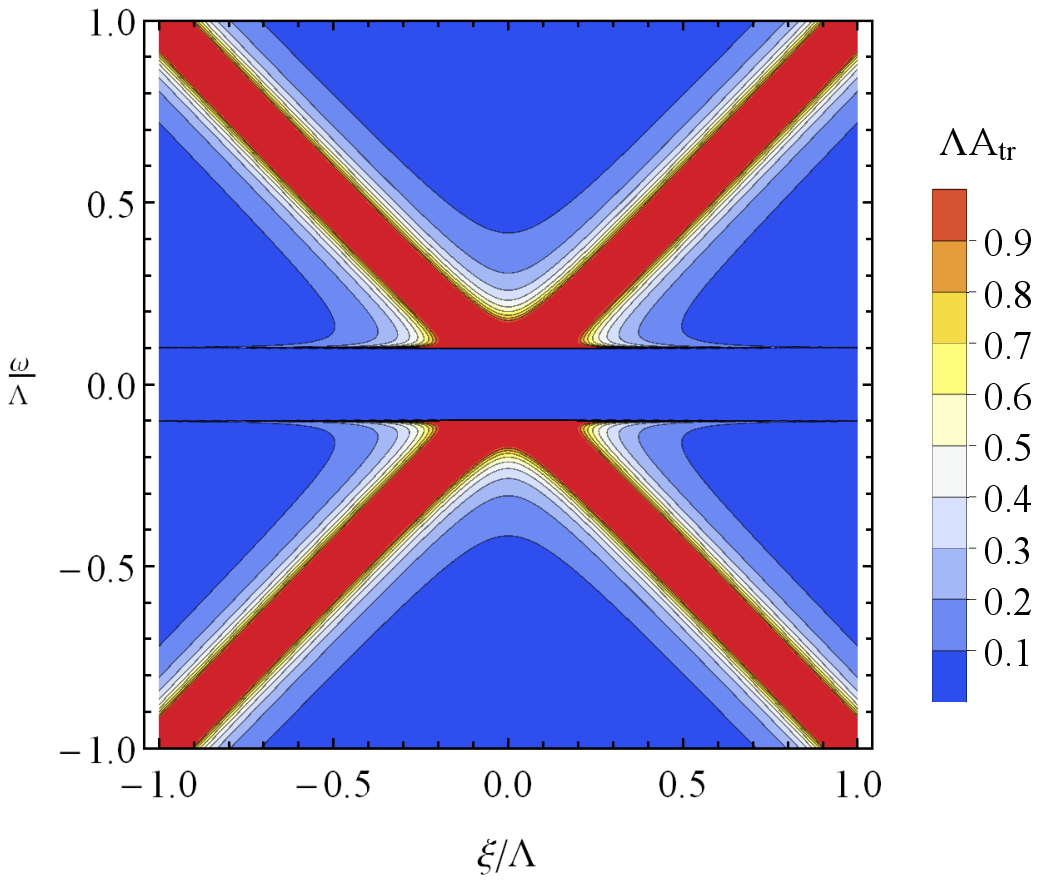}
\end{center}
\hspace{0.075\textwidth}{\small (a)}\hspace{0.525\textwidth}{\small (b)}\\[0pt]
\caption{The trace of the spectral function $A_{\rm tr}=\mbox{tr}A(\omega,\mathbf{p})$ as a function of $\omega$ and $\xi$ for $\Delta=0$ (panel (a)) as well as both  $\Delta=\alpha$ and $\Delta=\alpha \sign{\omega}$ (panel (b)). In both panels we set $\tau=10/\Lambda$, $\alpha=0.1\Lambda$, and $T=0$.}
\label{fig:Disorder-Atr-s-wave-7-6-8}
\end{figure*}

In order to find characteristic spectroscopic signatures of the Berezinskii pairing, let us consider the simplest case of s-wave pairing. We present the trace of the spectral function $A_{\rm tr}=\mbox{tr}A(\omega,\mathbf{p})$ as a function of $\omega$ and $\xi$ for $\Delta=0$, $\Delta=\alpha$, and $\Delta=\alpha \sign{\omega}$ in Fig.~\ref{fig:Disorder-Atr-s-wave-7-6-8}.
While the conventional superconducting gap $\Delta=\alpha$ has a well-defined qualitative manifestation in the spectral function, it is indistinguishable from the simplest OF gap $\Delta=\alpha \sign{\omega}$. Indeed, this follows from the fact that the trace of the spectral function depends only on the absolute value of the gap.

\begin{figure*}[!ht]
\begin{center}
\includegraphics[width=0.32\textwidth]{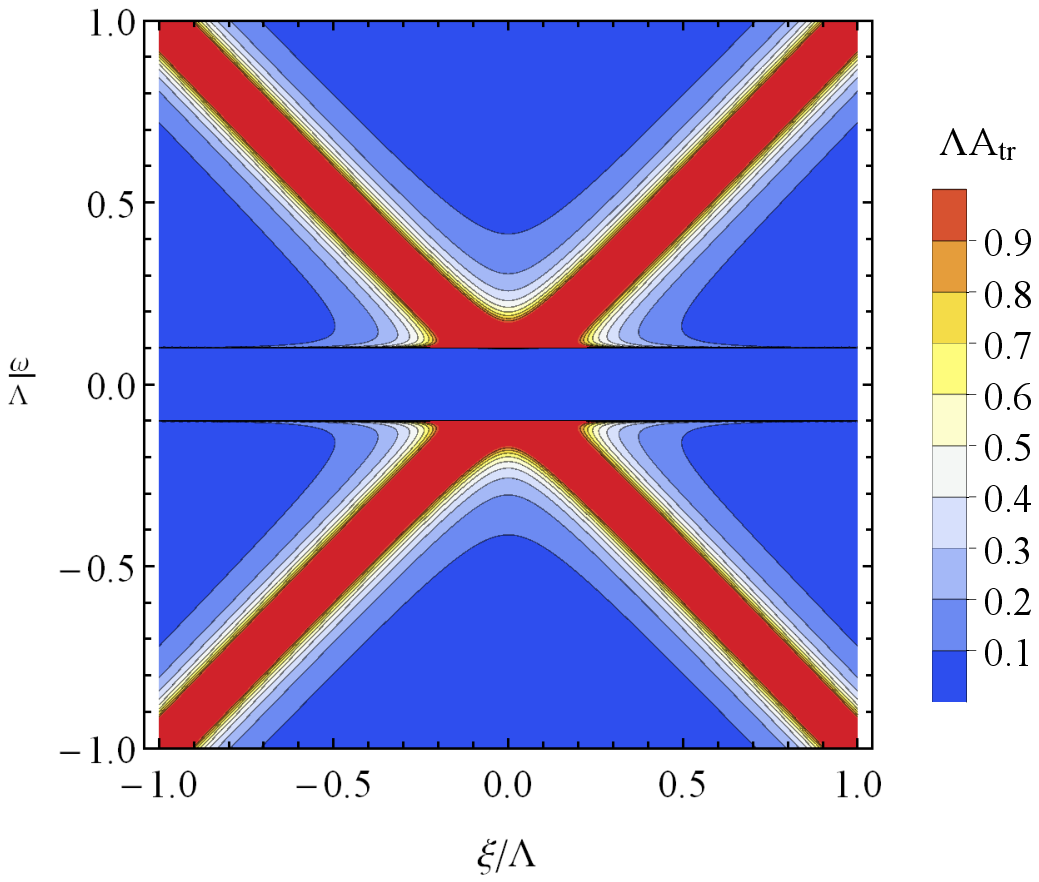}\hfill
\includegraphics[width=0.32\textwidth]{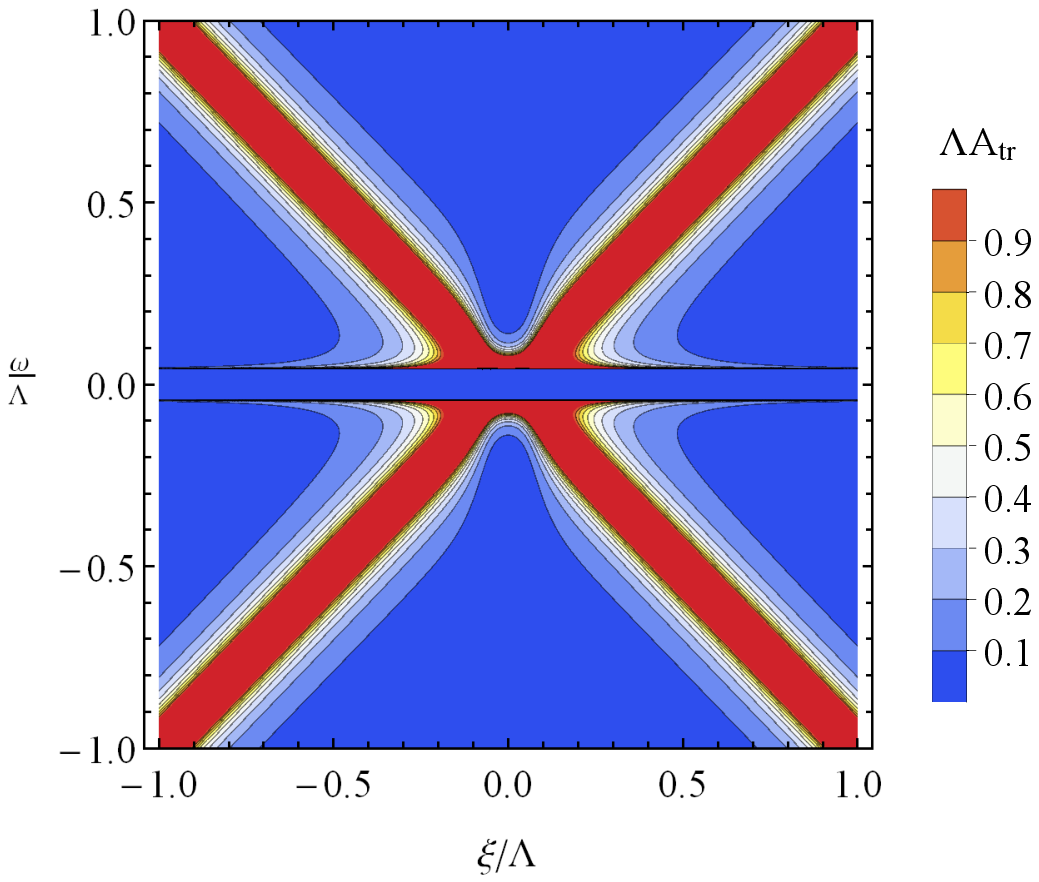}\hfill
\includegraphics[width=0.32\textwidth]{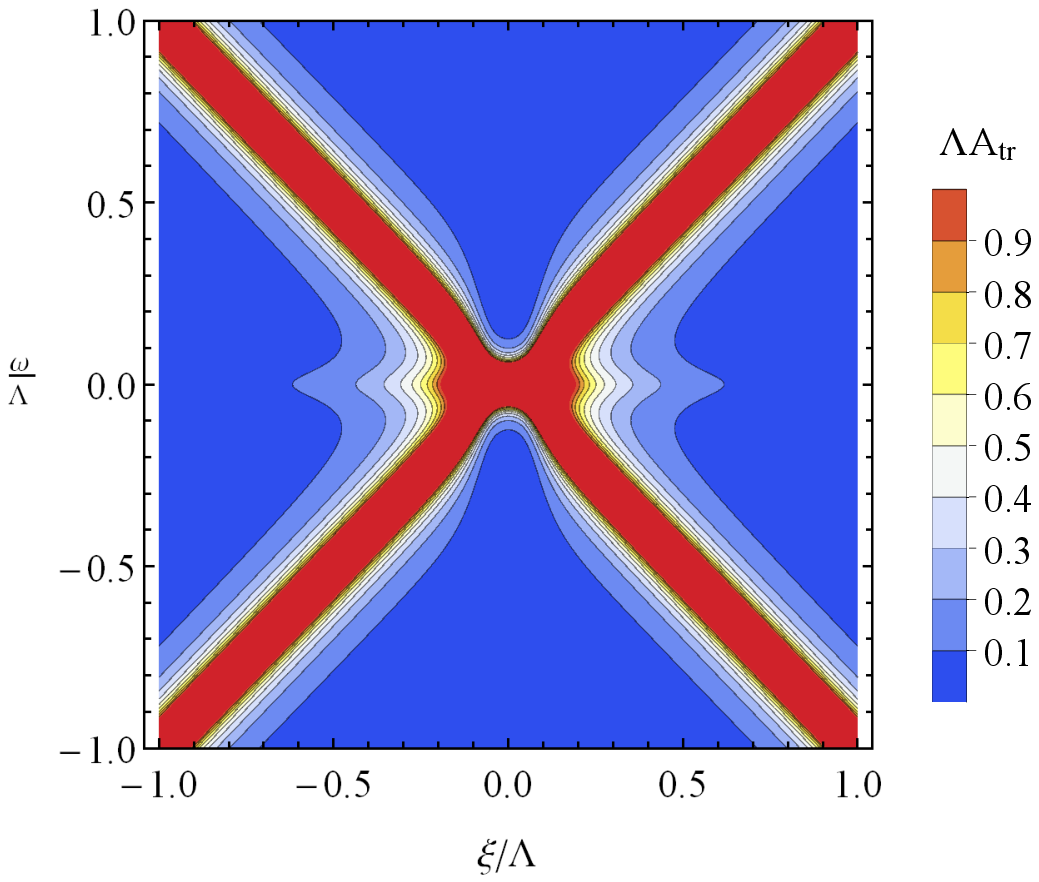}
\end{center}
\hspace{0.01\textwidth}{\small (a)}\hspace{0.3\textwidth}{\small (b)}\hspace{0.3\textwidth}{\small (c)}\\[0pt]
\caption{The trace of the spectral function $A_{\rm tr}=\mbox{tr}A(\omega,\mathbf{p})$ for $\Delta=\alpha \omega/\sqrt{\omega^2 +\beta^2 \Lambda^2}$ at $\beta=0.01$ (panel (a)), $\beta=0.09$ (panel (b)), and $\beta=0.11$ (panel (c)). We set $\tau=10/\Lambda$, $T=0$, and $\alpha=0.1\Lambda$.}
\label{fig:Disorder-Atr-s-wave-4}
\end{figure*}

\begin{figure*}[!ht]
\begin{center}
\includegraphics[width=0.32\textwidth]{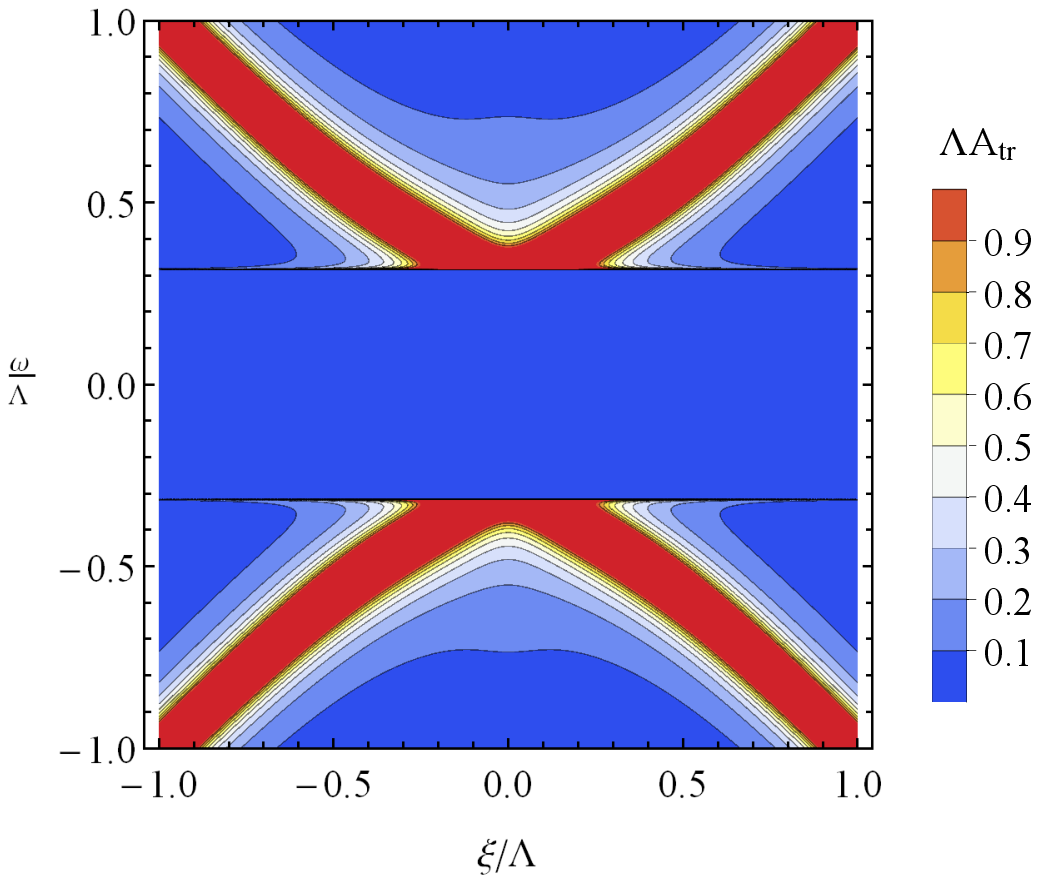}\hfill
\includegraphics[width=0.32\textwidth]{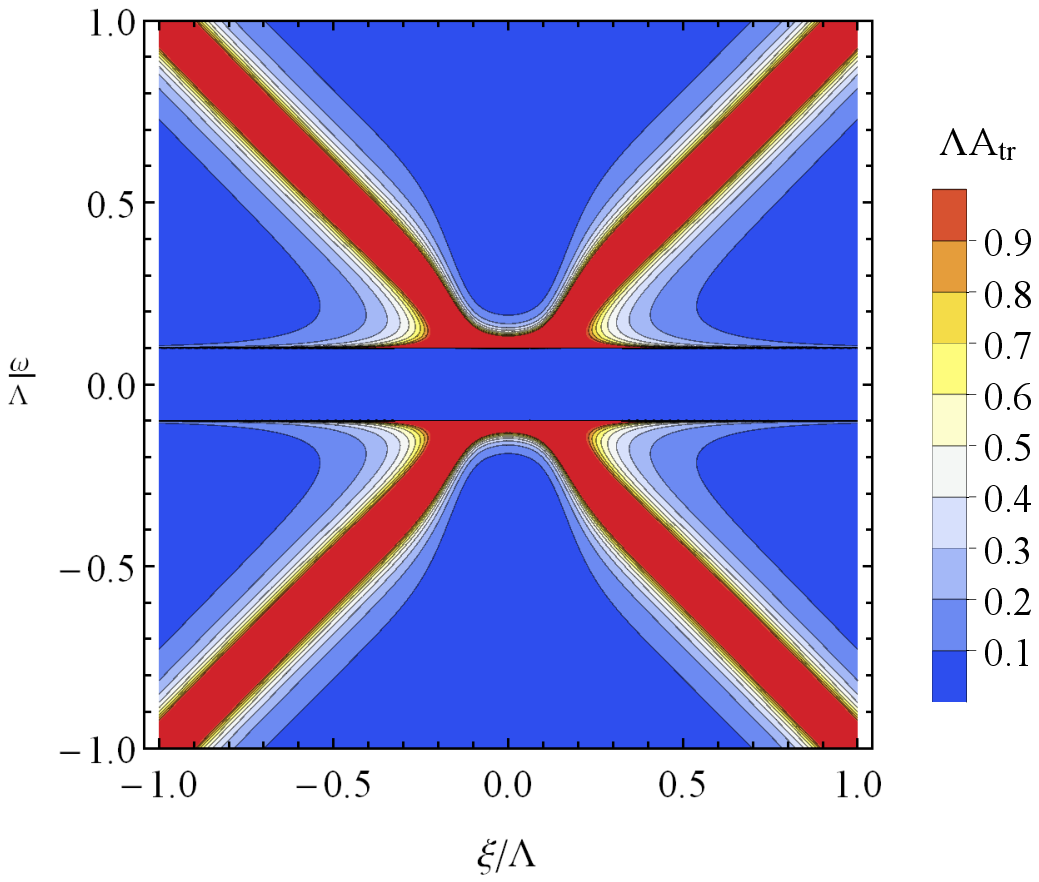}\hfill
\includegraphics[width=0.32\textwidth]{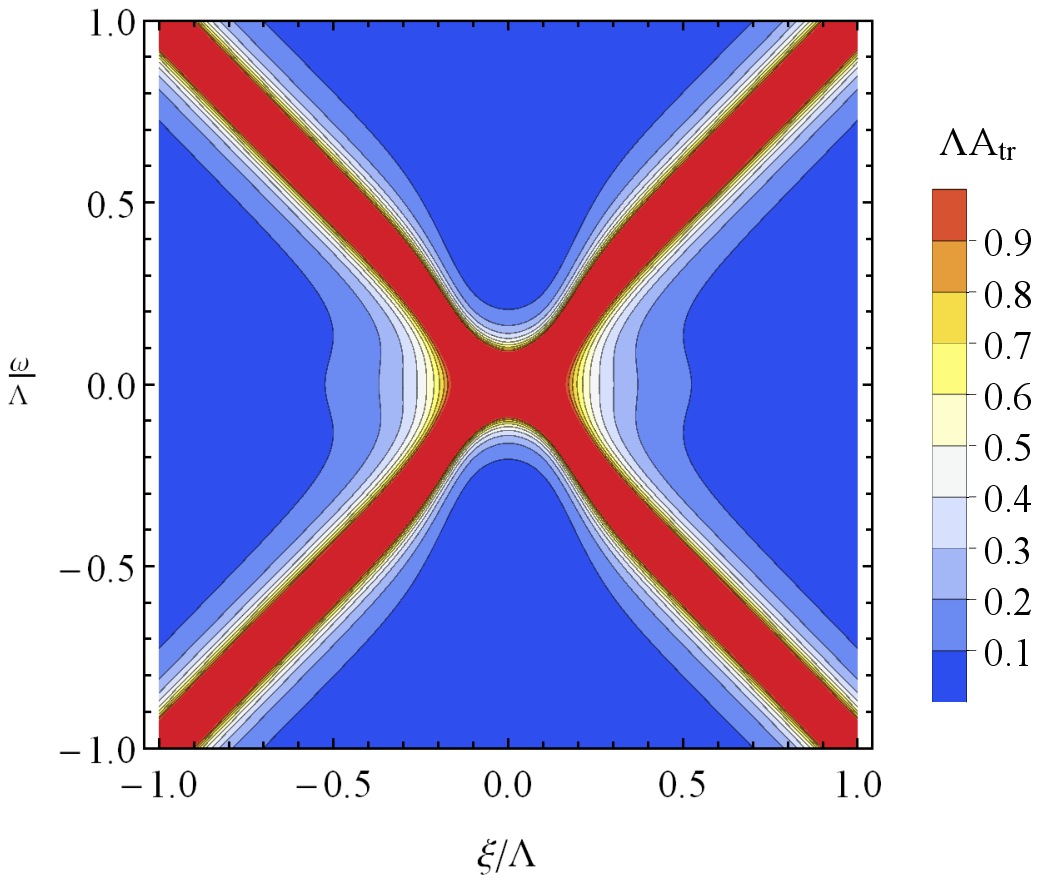}
\end{center}
\hspace{0.01\textwidth}{\small (a)}\hspace{0.3\textwidth}{\small (b)}\hspace{0.3\textwidth}{\small (c)}\\[0pt]
\caption{The trace of the spectral function $A_{\rm tr}=\mbox{tr}A(\omega,\mathbf{p})$ for $\Delta=\alpha \omega \Lambda/\left(\omega^2 +\beta^2 \Lambda^2\right)$ at $\beta=0.01$ (panel (a)), $\beta=0.3$ (panel (b)), and $\beta=0.35$ (panel (c)). We set $\tau=10/\Lambda$, $T=0$, and $\alpha=0.1\Lambda$.}
\label{fig:Disorder-Atr-s-wave-2}
\end{figure*}

The results for the OF gaps $\Delta=\alpha \omega/\sqrt{\omega^2 +\beta^2 \Lambda^2}$ and $\Delta=\alpha \omega \Lambda/\left(\omega^2 +\beta^2 \Lambda^2\right)$ are shown in Figs.~\ref{fig:Disorder-Atr-s-wave-4} and \ref{fig:Disorder-Atr-s-wave-2}, respectively. As expected, $A_{\rm tr}$ for the small values of $\beta$ in Fig.~\ref{fig:Disorder-Atr-s-wave-4}(a) almost coincides with that for both $\Delta=\alpha$ and $\Delta=\alpha \sign{\omega}$ in Fig.~\ref{fig:Disorder-Atr-s-wave-7-6-8}(b). With the increase of $\beta$, the gap in the spectrum slowly diminishes until it completely closes at $\beta\approx\alpha/\Lambda$. Then, the spectral function quickly reduces to that in the normal state and at large $\beta$ is almost undistinguishable from the results in Fig.~\ref{fig:Disorder-Atr-s-wave-7-6-8}(a).
The dependence of $A_{\rm tr}$ on $\omega$ and $\xi$ for $\Delta=\alpha \omega \Lambda/\left(\omega^2 +\beta^2 \Lambda^2\right)$ is similar and is presented in Fig.~\ref{fig:Disorder-Atr-s-wave-2}. While the gap is clearly visible at small, it disappears at large values of $\beta$. For both ansatzes, the results at large $\beta$ are almost indistinguishable from those for the normal state in Fig.~\ref{fig:Disorder-Atr-s-wave-7-6-8}(a).

In order to calculate the critical values of the parameter $\beta$ that corresponds to the vanishing of the spectral gap, let us consider the frequencies around which the spectral function is peaked. In the clean case, they are determined by the following equations for the gap ansatzes $\Delta=\alpha \omega/\sqrt{\omega^2 +\beta^2 \Lambda^2}$ and $\Delta=\alpha \omega \Lambda/\left(\omega^2 +\beta^2 \Lambda^2\right)$, respectively:
\begin{eqnarray}
\label{Disorder-poles-1}
&&\omega^2\frac{\beta^2\Lambda^2 -\alpha^2 +\omega^2}{\omega^2 +\beta^2 \Lambda^2}=\xi^2,\\
\label{Disorder-poles-2}
&&\omega^2\frac{(\omega^2+\beta^2\Lambda^2)^2 -\alpha^2\Lambda^2}{(\omega^2 +\beta^2 \Lambda^2)^2}=\xi^2.
\end{eqnarray}
It is clear that at $|\omega|\ll\Lambda, \alpha$ the roots of the Eqs.~(\ref{Disorder-poles-1}) and (\ref{Disorder-poles-2}) are absent for $|\beta\Lambda|<|\alpha|$ and $|\beta^2\Lambda|<|\alpha|$, respectively. Therefore, the critical values of $\beta$ at which the spectral gap closes are $\beta=\alpha/\Lambda$ for $\Delta=\alpha \omega/\sqrt{\omega^2 +\beta^2 \Lambda^2}$ and $\beta=\sqrt{\alpha/\lambda}$ for $\Delta=\alpha \omega \Lambda/\left(\omega^2 +\beta^2 \Lambda^2\right)$.

\subsection{Density of states}
\label{sec:DOS-D}

The electron density of states is another important quantity that is used to probe the formation of the superconducting gap. In the model at hand, it reads as
\begin{widetext}
\begin{eqnarray}
\label{Disorder-DOS-e-def}
\nu(\omega) &=& -\frac{1}{\pi} \mbox{Im}\int\frac{d^np}{(2\pi)^n} \mbox{tr}\left[\frac{1+\tau_z}{2}\hat{G}^{\rm R}(\omega,\mathbf{p})\right]
= -\frac{2}{\pi} \frac{\nu_0}{4\pi} \int d\Omega_{\mathbf{p}} \int_{-\Lambda}^{\Lambda} d\xi \frac{1}{Z(\omega, \mathbf{p})}\nonumber\\
&\times& \frac{\eta_{\rm Im} \omega \left\{\left[\omega^2 - |\Delta(\omega,\mathbf{p})|^2\right] \left[\eta_{\rm Re}^2 - \eta_{\rm Im}^2\right] -\xi^2/Z^2(\omega, \mathbf{p})\right\} -2\eta_{\rm Re}^2 \eta_{\rm Im} \left[\omega +\xi/Z(\omega, \mathbf{p})\right] \left[\omega^2 -|\Delta(\omega,\mathbf{p})|^2\right]}{\left\{\left[\omega^2 -|\Delta(\omega,\mathbf{p})|^2\right] \left(\eta_{\rm Re}^2 - \eta_{\rm Im}^2\right) -\xi^2/Z^2(\omega, \mathbf{p})\right\}^2 + \left\{2\eta_{\rm Re}\eta_{\rm Im}\left[\omega^2 -|\Delta(\omega,\mathbf{p})|^2\right]\right\}^2}.
\end{eqnarray}
\end{widetext}

\begin{figure*}[!ht]
\begin{center}
\includegraphics[width=0.45\textwidth]{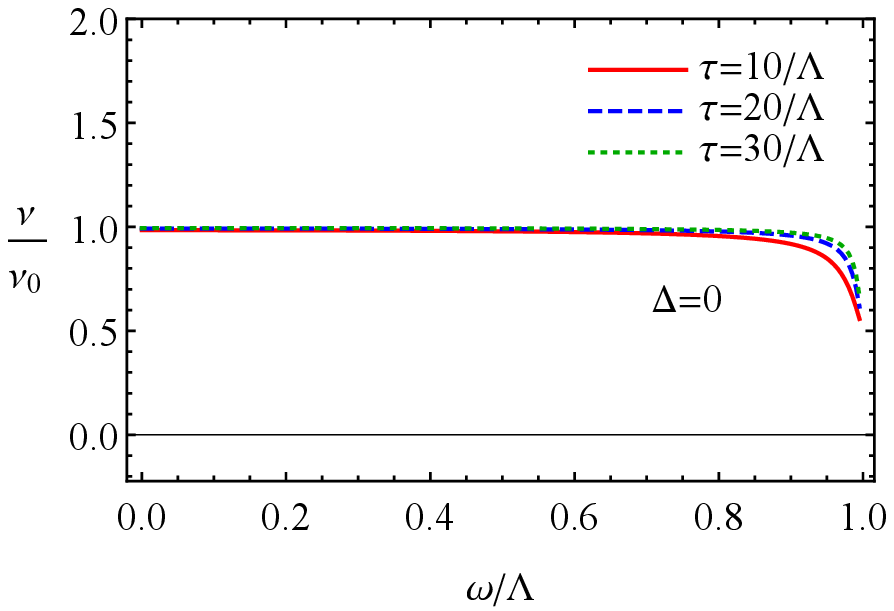}\hfill
\includegraphics[width=0.45\textwidth]{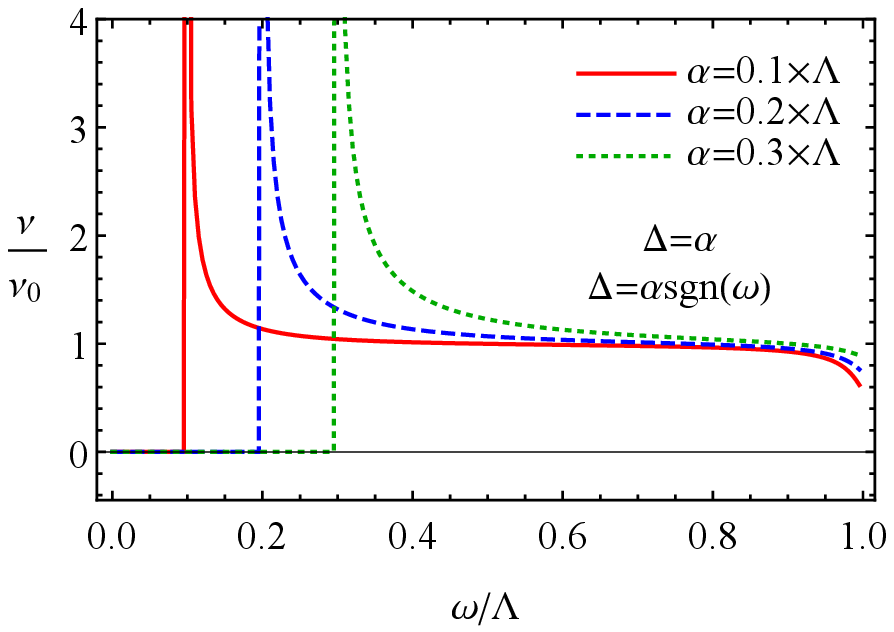}
\end{center}
\hspace{0.075\textwidth}{\small (a)}\hspace{0.525\textwidth}{\small (b)}\\[0pt]
\caption{The electron DOS $\nu$ as a function of $\omega$ for $\Delta=0$ (panel (a)) at $\alpha=0.1\Lambda$ and a few values of $\tau$ as well as $\Delta=\alpha$ and $\Delta=\alpha \sign{\omega}$ (panel (b)) at $\tau=10/\Lambda$ and a few values of $\alpha$. We set $T=0$ in both panels.}
\label{fig:Disorder-DOS-s-wave-7-6-8}
\end{figure*}

The results for the DOS in the case of $\Delta=0$, $\Delta=\alpha$, and $\Delta=\alpha \sign{\omega}$ are shown in Fig.~\ref{fig:Disorder-DOS-s-wave-7-6-8}.
For definiteness, we assumed s-wave pairing in 2D. Note, however, that due to the approximation of a large Fermi surface, the results are qualitatively the same for 3D superconductors too. The deviations from $\nu_0$ at large $\omega$ in Fig.~\ref{fig:Disorder-DOS-s-wave-7-6-8}(a) are related to the disorder effect and the use of a finite cutoff $\Lambda$ in the integration over $\xi$. As expected, the coherence peaks at $|\omega|=|\alpha|$ are formed for the BCS type of the gap, $\Delta=\alpha$ (see Fig.~\ref{fig:Disorder-DOS-s-wave-7-6-8}(b)). Since the DOS (\ref{Disorder-DOS-e-def}) is sensitive only to the absolute value of the gap, the same dependence is also valid for the OF gap $\Delta=\alpha \sign{\omega}$.

\begin{figure*}[!ht]
\begin{center}
\includegraphics[width=0.45\textwidth]{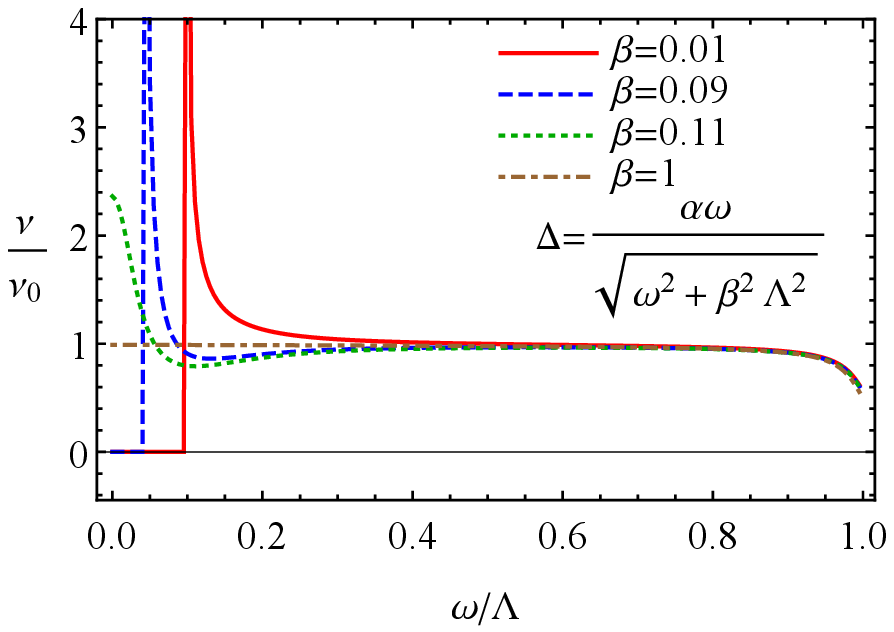}\hfill
\includegraphics[width=0.45\textwidth]{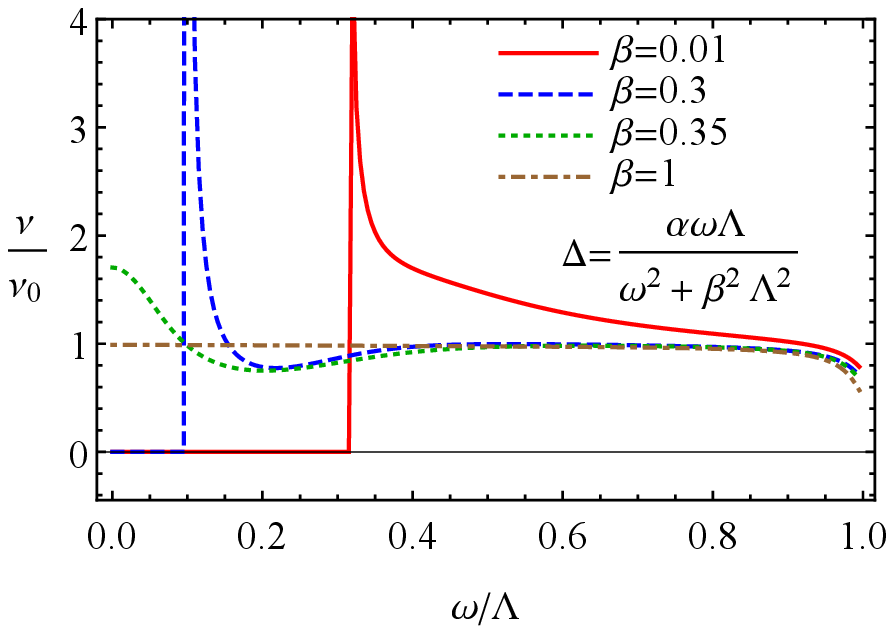}
\end{center}
\hspace{0.075\textwidth}{\small (a)}\hspace{0.525\textwidth}{\small (b)}\\[0pt]
\caption{The electron DOS $\nu$ as a function of $\omega$ for $\Delta=\alpha \omega/\sqrt{\omega^2 +\beta^2 \Lambda^2}$ (panel (a)) and $\Delta=\alpha \omega \Lambda/\left(\omega^2 +\beta^2 \Lambda^2\right)$ (panel (b)). We set $\tau=10/\Lambda$, $\alpha=0.1\Lambda$, and $T=0$ in both panels.}
\label{fig:Disorder-DOS-s-wave-4-2}
\end{figure*}

The electron DOS as a function of $\omega$ for the OF gaps $\Delta=\alpha \omega/\sqrt{\omega^2 +\beta^2 \Lambda^2}$ and $\Delta=\alpha \omega \Lambda/\left(\omega^2 +\beta^2 \Lambda^2\right)$ is presented in Figs.~\ref{fig:Disorder-DOS-s-wave-4-2}(a) and \ref{fig:Disorder-DOS-s-wave-4-2}(b), respectively, for a few values of $\beta$. In agreement with the results for the spectral functions presented in Figs.~\ref{fig:Disorder-Atr-s-wave-4} and \ref{fig:Disorder-Atr-s-wave-2}, the DOS evolves from the BCS-like form with the characteristic coherence peak at small values of $\beta$ to the normal-like form with an almost constant DOS at large $\beta$. The results for the intermediate values of $\beta$ are, however, nontrivial. As one can see in Fig.~\ref{fig:Disorder-DOS-s-wave-4-2}, the gap in the DOS closes at the critical values of $\beta$ discussed after Eqs.~(\ref{Disorder-poles-1}) and (\ref{Disorder-poles-2}) leading to an enhancement of the DOS at $\omega=0$. This feature quickly diminish with the flattening of the frequency profile of the gap, however. In addition, the electron DOS shows a well pronounced local minimum while approaching the peak for intermediate values of $\beta$. The origin of this feature is related to the coefficient $Z(\xi)$ and the conservation of the spectral density.

Our results suggest that the trace of the spectral function $A_{\rm tr}=\mbox{tr}A(\omega,\mathbf{p})$ and the electron DOS $\nu$ demonstrate a few interesting features. For example, unlike naive expectations, the Berezinskii pairing might lead to the spectral gap despite the fact that $\Delta(\omega)$ is odd in frequency. Indeed, this follows from the self-consistent solution for the poles of the Green function, which are absent for small values of $\omega$. Further, depending on the frequency profiles of the gap ansatzes, the spectral gap might close when the frequency dependence is weak.
In general, however, it might be hard to distinguish the OF gaps from their even-frequency counterparts by studying the spectral function and the electron DOS only.
The corresponding distinctive features of the Berezinskii paring might appear in the optical conductivity, which will be discussed in the next section.

\section{Optical conductivity}
\label{sec:OC-calc}

\subsection{General definitions}
\label{sec:OC-calc-def}

The optical conductivity tensor is given by the following standard expression in the Kubo linear response formalism (see, e.g., Refs.~\cite{Bruus-book,Altland:book,Mahan:book}):
\begin{eqnarray}
\label{Disorder-sigma-def}
\sigma_{ij}(\Omega) = \frac{i}{2\Omega} \left\{\frac{e^2}{m} \delta_{ij}\left\langle \hat{G}\left(0, \mathbf{0}\right)\right\rangle +\Pi_{ij}\left(\Omega+i0, \mathbf{0}\right)\right\}.
\end{eqnarray}
Here the first term in the curly brackets $\propto\left\langle \hat{G}\left(0, \mathbf{0}\right)\right\rangle$ is the diamagnetic term and the second one, which is determined by the polarization operator $\propto\Pi_{ij}\left(\Omega+i0, \mathbf{0}\right)$, is the paramagnetic term. By definition, $\left\langle \hat{G}\left(0, \mathbf{0}\right)\right\rangle=2n$. [An additional factor $2$ comes from the fact that the Green function in the Nambu space $\hat{G}\left(0, \mathbf{0}\right)$ effectively contains doubled number of the degrees of freedom.] Here the fermion number density is
\begin{eqnarray}
\label{Disorder-sigma-n-def}
\mbox{2D:} \quad n &=& \frac{p_{\rm F}^2}{4\pi},\\
\mbox{3D:} \quad n &=& \frac{p_{\rm F}^3}{6\pi^2},
\end{eqnarray}
and $p_{\rm F}=\sqrt{2m\mu}$ is the Fermi momentum.

The general expression for the polarization tensor reads as
\begin{eqnarray}
\label{Disorder-Pi-def}
&&\Pi_{ij}\left(\Omega+i0, \mathbf{0}\right) = T \sum_{l=-\infty}^{\infty} \int\frac{d^np}{(2\pi)^n} \nonumber\\ &&\times\mbox{tr}\left[\hat{j}_{i}\hat{G}\left(i\omega_l,\mathbf{p}\right)\hat{j}_{j} \hat{G}\left(i\omega_l-\Omega-i0,\mathbf{p}\right)\right].
\end{eqnarray}
Here $\omega_l=(2l+1)\pi T$ are the fermionic Matsubara frequencies, $T$ is temperature, $\hat{\mathbf{j}} = -e\mathbf{p}/m \mathds{1}_2\approx -e p_{\rm F} \hat{\mathbf{p}}/m \mathds{1}_2$ is the electric current operator, and $-e$ is the electron charge. By using the relation for the Green function (\ref{Disorder-SP-G-A}), one can rewrite Eq.~(\ref{Disorder-Pi-def}) in a more convenient form, i.e.,
\begin{eqnarray}
\label{Disorder-Pi-1}
&&\Pi_{ij}\left(\Omega+i0, \mathbf{0}\right) = T \sum_{l=-\infty}^{\infty} \int d\omega \int d\omega^{\prime} \int\frac{d^np}{(2\pi)^n} \nonumber\\
&&\times\frac{1}{i\omega_l-\omega} \frac{1}{i\omega_l-\omega^{\prime}-\Omega-i0} \mbox{tr}\left[\hat{j}_{i}\hat{A}\left(\omega,\mathbf{p}\right)\hat{j}_{j}\hat{A}\left(\omega^{\prime},\mathbf{p}\right)\right] \nonumber\\
&&= \int d\omega \int d\omega^{\prime} \frac{\nu_0}{4\pi} \int d\Omega_{\mathbf{p}} \int_{-\Lambda}^{\Lambda}d\xi \frac{n_{F}(\omega)-n_{F}(\omega^{\prime})}{\omega-\omega^{\prime}-\Omega-i0} \nonumber\\
&&\times\mbox{tr}\left[\hat{j}_{i}\hat{A} \left(\omega,\mathbf{p}\right)\hat{j}_{j}\hat{A}\left(\omega^{\prime},\mathbf{p}\right)\right].
\end{eqnarray}
Here the summation over the Matsubara frequencies was performed in the second equality and $n_{F}(\omega)=1/\left(1+e^{\omega/T}\right)$ is the standard Fermi--Dirac distribution function.

Let us start with the real part of the conductivity tensor $\mbox{Re}\,\sigma_{ij}(\Omega)$. As follows form Eq.~(\ref{Disorder-sigma-def}), it is determined by the imaginary part of the polarization tensor, which reads as
\begin{eqnarray}
\label{Disorder-Pi-Re}
&&\mbox{Im}\,\Pi_{ij}\left(\Omega+i0, \mathbf{0}\right) = \frac{e^2\pi p_{\rm F}^2 \nu_0}{m^2} \int d\omega \int \frac{d\Omega_{\mathbf{p}}}{4\pi} \int_{-\Lambda}^{\Lambda}d\xi \nonumber\\
&&\times\left[n_{F}(\omega)-n_{F}(\omega-\Omega)\right] \mbox{tr}\left[\hat{p}_{i}\hat{A} \left(\omega,\mathbf{p}\right)\hat{p}_{j}\hat{A}\left(\omega-\Omega,\mathbf{p}\right)\right].\nonumber\\
\end{eqnarray}
Here we used the Sokhotski formula
\begin{equation}
\label{Disorder-Sokhotski}
\frac{1}{x \pm i0} = \mbox{v.p.}\frac{1}{x} \mp i \pi \delta(x),
\end{equation}
where $\mbox{v.p.}$ stands for the principal value.

Thus, the real part of the electric conductivity tensor reads
\begin{eqnarray}
\label{Disorder-sigma-Re}
&&\mbox{Re}\,\sigma_{ij}(\Omega) = -\frac{e^2\pi p_{\rm F}^2 \nu_0}{2\Omega m^2} \int d\omega \int \frac{d\Omega_{\mathbf{p}}}{4\pi} \int_{-\Lambda}^{\Lambda}d\xi \nonumber\\
&&\times \left[n_{F}(\omega)-n_{F}(\omega-\Omega)\right] \mbox{tr}\left[\hat{p}_{i}\hat{A} \left(\omega,\mathbf{p}\right)\hat{p}_{j}\hat{A}\left(\omega-\Omega,\mathbf{p}\right)\right],\nonumber\\
\end{eqnarray}
where the explicit expression for the spectral function $\hat{A} \left(\omega,\mathbf{p}\right)$ is given in Eq.~(\ref{Disorder-SP-A}).

It is convenient to normalize the optical conductivity by the dc conductivity in the normal phase, which reads as
\begin{eqnarray}
\label{Disorder-sigma-0-def}
\sigma_0 = \frac{ne^2 \tau}{m}.
\end{eqnarray}

The imaginary part of the optical conductivity tensor $\sigma_{ij}(\Omega)$ consists of two parts: (i) the diamagnetic term quantified by $\left\langle G\left(0, \mathbf{0}\right)\right\rangle$ in Eq.~(\ref{Disorder-sigma-def}) and (ii) the paramagnetic term determined by the polarization operator $\Pi_{ij}\left(\Omega+i0, \mathbf{0}\right)$. The diamagnetic term immediately follows from the definition of the density $\left\langle \hat{G}\left(0, \mathbf{0}\right)\right\rangle=2n$ and reads as
\begin{eqnarray}
\label{Disorder-sigma-Im-diam}
\mbox{Im}\,\sigma_{ij}^{\rm diam}(\Omega) = \frac{e^2 n}{\Omega m} \delta_{ij}.
\end{eqnarray}
The paramagnetic contribution is
\begin{eqnarray}
\label{Disorder-sigma-Im-param}
&&\mbox{Im}\,\sigma_{ij}^{\rm param}(\Omega) = \frac{e^2p_{\rm F}^2\nu_0}{2\Omega m^2} \mbox{v.p.}\int d\omega^{\prime} \int d\omega \int \frac{d\Omega_{\mathbf{p}}}{4\pi} \nonumber\\
&&\times\int_{-\Lambda}^{\Lambda}d\xi \frac{n_{F}(\omega)-n_{F}(\omega^{\prime})}{\omega-\omega^{\prime}-\Omega} \mbox{tr}\left[\hat{p}_{i}\hat{A} \left(\omega,\mathbf{p}\right)\hat{p}_{j}\hat{A}\left(\omega^{\prime},\mathbf{p}\right)\right],\nonumber\\
\end{eqnarray}
where Eq.~(\ref{Disorder-Sokhotski}) was used. Thus, the imaginary part of the conductivity reads as
\begin{eqnarray}
\label{Disorder-sigma-Im-total}
&&\mbox{Im}\,\sigma_{ij}(\Omega) = \frac{e^2 n}{\Omega m}  \delta_{ij} + \frac{e^2p_{\rm F}^2\nu_0}{2\Omega m^2} \mbox{v.p.}\int d\omega^{\prime} \int d\omega \int \frac{d\Omega_{\mathbf{p}}}{4\pi} \nonumber\\
&&\times\int_{-\Lambda}^{\Lambda}d\xi \frac{n_{F}(\omega)-n_{F}(\omega^{\prime})}{\omega-\omega^{\prime}-\Omega} \mbox{tr}\left[\hat{p}_{i}\hat{A} \left(\omega,\mathbf{p}\right)\hat{p}_{j}\hat{A}\left(\omega^{\prime},\mathbf{p}\right)\right].\nonumber\\
\end{eqnarray}

In order to simplify the calculations, it is convenient to consider the clean limit $\tau \to \infty$ for the imaginary part of the conductivity tensor. Then, Eq.~(\ref{Disorder-sigma-Im-total}) can be rewritten as
\begin{widetext}
\begin{eqnarray}
\label{Disorder-sigma-Im-clean}
\mbox{Im}\,\sigma_{ij}(\Omega) &=& \frac{e^2 n}{\Omega m}  \delta_{ij} +\frac{e^2p_{\rm F}^2\nu_0}{2\Omega m^2} \mbox{v.p.}\int d\omega^{\prime} \int d\omega \int \frac{d\Omega_{\mathbf{p}}}{4\pi} \int_{-\Lambda}^{\Lambda}d\xi \frac{n_{F}(\omega)-n_{F}(\omega^{\prime})}{\omega-\omega^{\prime}-\Omega} \hat{p}_i\hat{p}_j \sign{\omega} \sign{\omega^{\prime}} \nonumber\\
&\times& \delta\left[Z^2(\omega, \mathbf{p})\omega^2-\xi^2 -Z^2(\omega, \mathbf{p})|\Delta(\omega,\mathbf{p})|^2\right] \delta\left[Z^2(\omega^{\prime}, \mathbf{p})(\omega^{\prime})^2 -\xi^2 -Z^2(\omega^{\prime}, \mathbf{p})|\Delta(\omega^{\prime},\mathbf{p})|^2\right] \nonumber\\
&\times& Z(\omega, \mathbf{p})Z(\omega^{\prime}, \mathbf{p}) \Big\{2\omega\omega^{\prime} +2\frac{\xi^2}{Z(\omega, \mathbf{p})Z(\omega^{\prime}, \mathbf{p})} +\Delta(\omega,\mathbf{p})\Delta^{\dag}(\omega^{\prime},\mathbf{p})  
+\Delta(\omega^{\prime},\mathbf{p})\Delta^{\dag}(\omega,\mathbf{p})\Big\}
=\frac{e^2 n}{\Omega m} \delta_{ij} \nonumber\\
&+&\frac{e^2p_{\rm F}^2\nu_0}{2\Omega m^2} \mbox{v.p.} \int \frac{d\Omega_{\mathbf{p}}}{4\pi} \int_{-\Lambda}^{\Lambda}d\xi \sum_{l,n} \frac{n_{F}(\tilde{\omega}_l)-n_{F}(\tilde{\omega}_n)}{\tilde{\omega}_l-\tilde{\omega}_n-\Omega} \hat{p}_i\hat{p}_j \sign{\tilde{\omega}_l} \sign{\tilde{\omega}_n} \nonumber\\
&\times& \frac{1}{\left|\partial_{\omega}\left[Z^2(\omega, \mathbf{p})\omega^2-\xi^2 -Z^2(\omega, \mathbf{p})|\Delta(\omega,\mathbf{p})|^2\right]\right|_{\omega\to \tilde{\omega}_{l}}}  \frac{1}{\left|\partial_{\omega^{\prime}}\left[Z^2(\omega^{\prime}, \mathbf{p})(\omega^{\prime})^2-\xi^2 -Z^2(\omega^{\prime}, \mathbf{p})|\Delta(\omega^{\prime},\mathbf{p})|^2\right]\right|_{\omega^{\prime}\to \tilde{\omega}_{n}}} \nonumber\\
&\times&Z(\tilde{\omega}_l, \mathbf{p})Z(\tilde{\omega}_n, \mathbf{p}) \Big\{2\tilde{\omega}_l\tilde{\omega}_n +2\frac{\xi^2}{Z(\tilde{\omega}_l, \mathbf{p})Z(\tilde{\omega}_n, \mathbf{p})} +\Delta(\tilde{\omega}_l,\mathbf{p})\Delta^{\dag}(\tilde{\omega}_n,\mathbf{p}) +\Delta(\tilde{\omega}_n,\mathbf{p})\Delta^{\dag}(\tilde{\omega}_l,\mathbf{p})\Big\},
\end{eqnarray}
\end{widetext}
where $\tilde{\omega}_{l}$ and $\tilde{\omega}_n$ are the real roots of equation $Z^2(\omega, \mathbf{p})\omega^2-\xi^2-Z^2(\omega, \mathbf{p})|\Delta(\omega,\mathbf{p})|^2=0$. The corresponding explicit expressions are cumbersome for the OF gaps $\Delta=\alpha \omega/\sqrt{\omega^2 +\beta^2 \Lambda^2}$ and $\Delta=\alpha \omega \Lambda/\left(\omega^2 +\beta^2 \Lambda^2\right)$. Therefore, we do not present them here.

In the following two subsection, the results for the real and imaginary parts of the optical conductivity will be discussed.

\subsection{Real part of the optical conductivity}
\label{sec:Disorder-results-Re}

Let us start our analysis with the real part of the optical conductivity. For simplicity, we assume that the pairing occurs in the s-wave channel. From the technical viewpoint, this case is much simpler since integrals over the angles are trivially performed. In addition, the conductivity tensor is diagonal, i.e., $\sigma_{ij}=\delta_{ij}\sigma$.

\begin{figure*}[!ht]
\begin{center}
\includegraphics[width=0.32\textwidth]{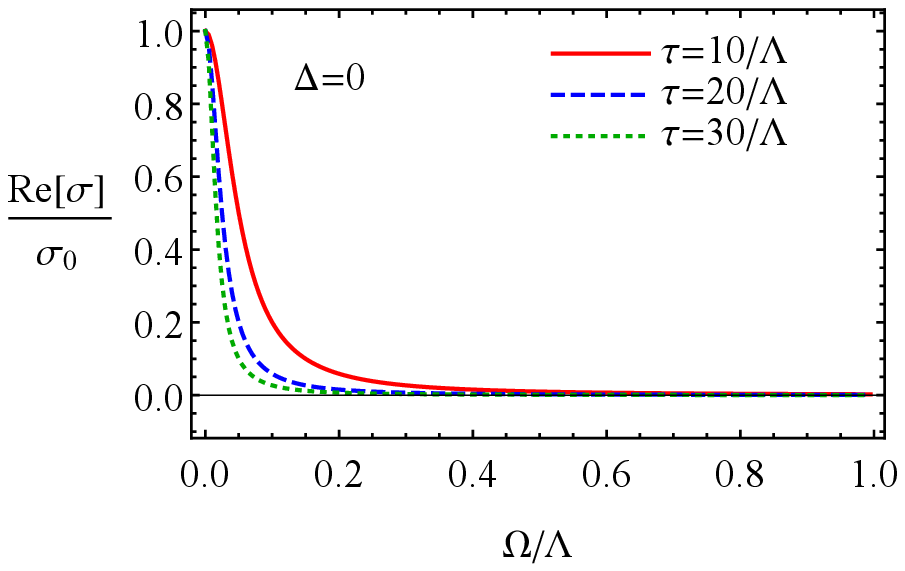}\hfill
\includegraphics[width=0.32\textwidth]{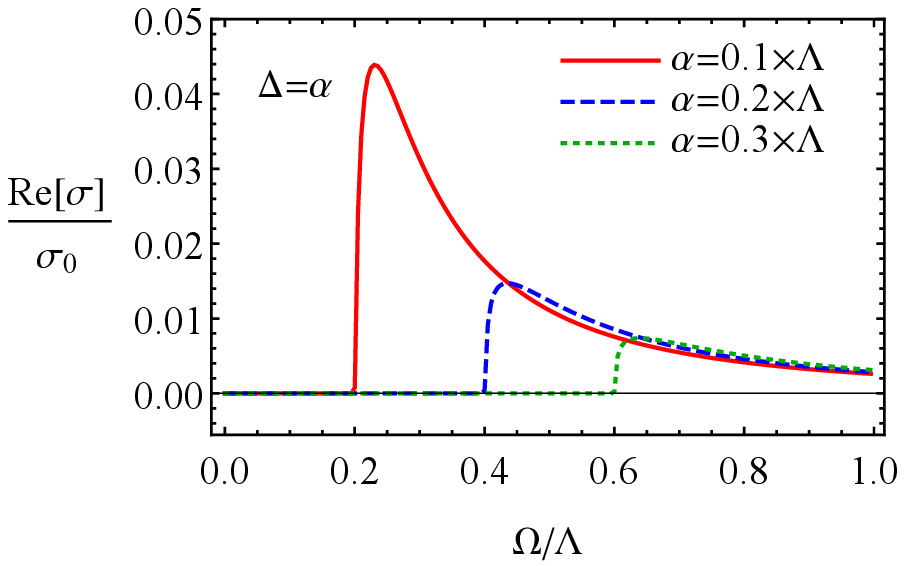}\hfill
\includegraphics[width=0.32\textwidth]{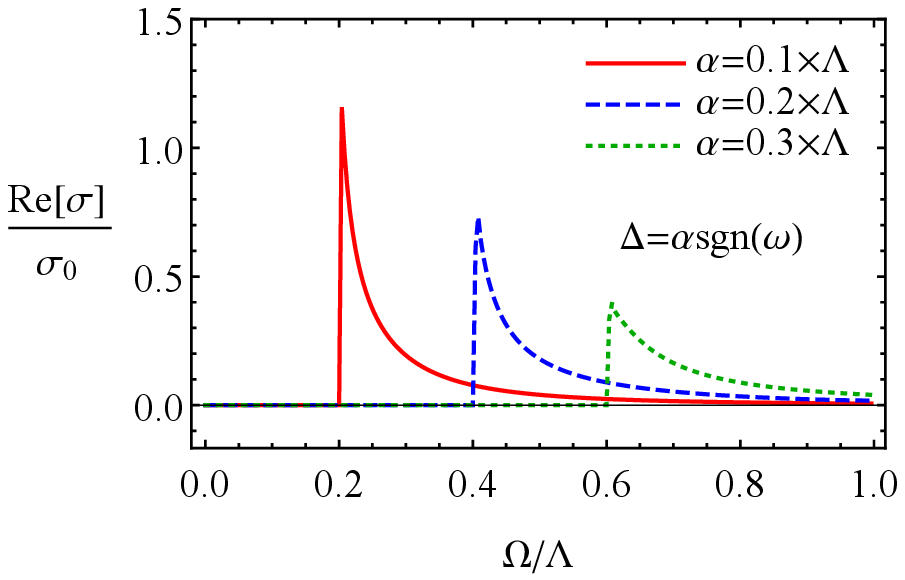}
\end{center}
\hspace{0.01\textwidth}{\small (a)}\hspace{0.3\textwidth}{\small (b)}\hspace{0.3\textwidth}{\small (c)}\\[0pt]
\caption{The real part of the optical conductivity for $\Delta=0$ at a few values of $\tau$ (panel (a)), as well as $\Delta=\alpha$ (panel (b)) and $\Delta=\alpha \sign{\omega}$ at a few values of $\alpha$ (panel (c)). We set $T=0$ in all panels and $\tau=10/\Lambda$ in panels (b) and (c).}
\label{fig:Disorder-results-s-wave-alpha=0}
\end{figure*}

The real part of the optical conductivity for $\Delta=0$ at a few $\tau$ and $\Delta=\alpha$ at a few $\alpha$ is given in Figs.~\ref{fig:Disorder-results-s-wave-alpha=0}(a) and \ref{fig:Disorder-results-s-wave-alpha=0}(b), respectively. As expected, the textbook Drude conductivity is reproduced at $\Delta=0$. Further, the results at $\Delta=\alpha$ suggest that the ac conductivity is zero at small frequencies $|\Omega|\lesssim 2|\Delta|$. The conductivity quickly increases at $|\Omega|\gtrsim 2|\Delta|$ and after reaching a peak slowly diminishes coinciding with $\sigma(\Omega)$ in the normal phase at $\Omega\to\infty$.
This finding agrees with the well-known result obtained by Mattis and Bardeen~\cite{Mattis-Bardeen:1958}, albeit includes the effects of non-magnetic disorder.

The peaks in the conductivity occur also for the Berezinskii pairing. In the case of a simple gap $\Delta(\omega) =\alpha \sign{\omega}$, they are shown in Fig.~\ref{fig:Disorder-results-s-wave-alpha=0}(c). While their positions are the same as for the BCS pairing shown in Fig.~\ref{fig:Disorder-results-s-wave-alpha=0}(b), the amplitude is significantly larger and the peaks are sharper. This observation might be useful to pinpoint the OF pairing in optical experiments. Moreover, the difference between the results in Figs.~\ref{fig:Disorder-results-s-wave-alpha=0}(b) and \ref{fig:Disorder-results-s-wave-alpha=0}(c) is particularly interesting since neither the spectral function nor the electron DOS can be used to distinguish the gaps $\Delta(\omega) =\alpha$ and $\Delta(\omega) =\alpha \sign{\omega}$. Indeed, while these spectroscopic methods depend only on the absolute value of the gap, the optical conductivity includes the terms $\propto \Delta (\omega)\Delta^{\dag}(\omega-\Omega)$, which is obviously different for $\Delta(\omega) =\alpha \sign{\omega}$ and $\Delta(\omega) =\alpha$.

\begin{figure*}[!ht]
\begin{center}
\includegraphics[width=0.32\textwidth]{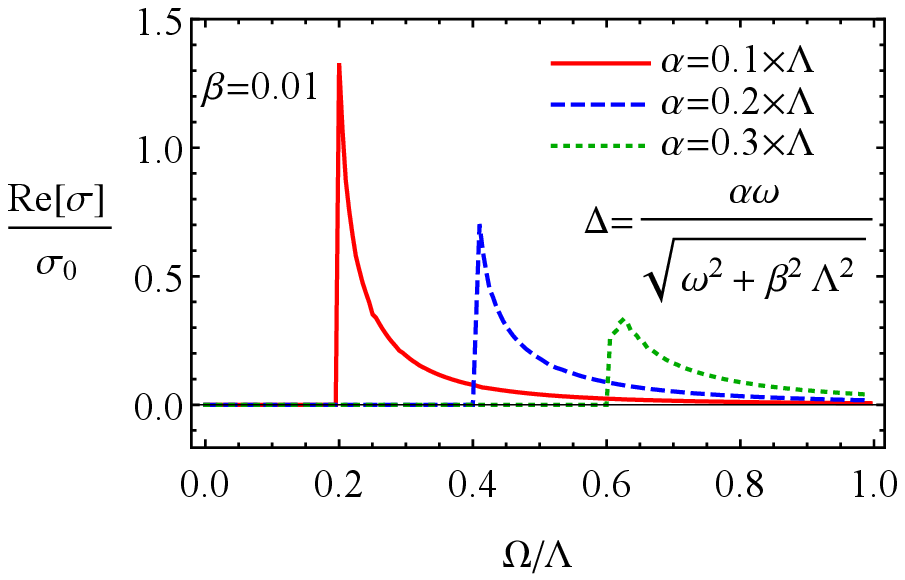}\hfill
\includegraphics[width=0.32\textwidth]{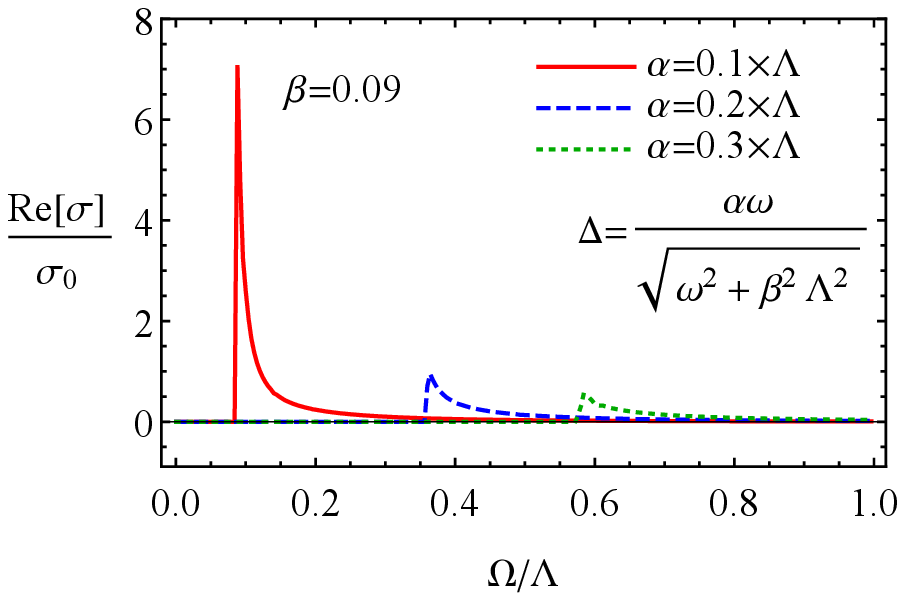}\hfill
\includegraphics[width=0.32\textwidth]{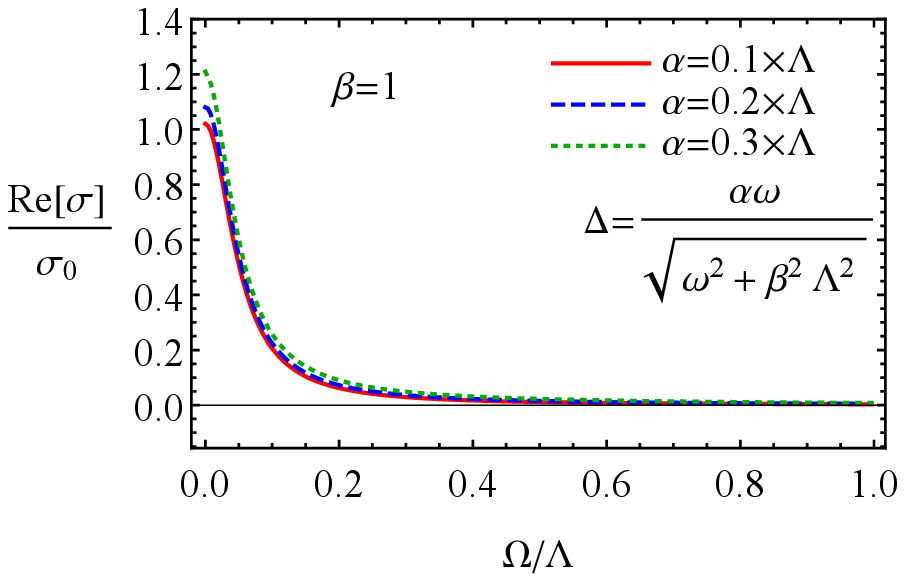}
\end{center}
\hspace{0.01\textwidth}{\small (a)}\hspace{0.3\textwidth}{\small (b)}\hspace{0.3\textwidth}{\small (c)}\\[0pt]
\caption{The real part of the optical conductivity for $\Delta= \alpha \omega/\sqrt{\omega^2 +\beta^2 \Lambda^2}$ at a few values of $\alpha$. We set $\beta=0.01$ (panel (a)), $\beta=0.09$ (panel (b)), and $\beta=1$ (panel (c)). In addition, $\tau=10/\Lambda$ and $T=0$ in all panels.
}
\label{fig:Disorder-results-s-wave-4}
\end{figure*}

In order to clarify what universal features exist for the Berezinskii pairing, let us consider more complicated ansatzes for the OF gaps, i.e., $\Delta=\alpha \omega/\sqrt{\omega^2 +\beta^2 \Lambda^2}$ and $\Delta=\alpha \omega \Lambda/\left(\omega^2 +\beta^2 \Lambda^2\right)$. The results for the former at three values of $\beta$ are shown in Fig.~\ref{fig:Disorder-results-s-wave-4}. As can be expected from the results for the spectral function and DOS in Sec.~\ref{sec:DOS}, the conductivity for small $\beta$ ($\beta=10^{-2}$) given in Fig.~\ref{fig:Disorder-results-s-wave-4}(a) is similar to that for $\Delta=\alpha \sign{\omega}$ and shows characteristic peaks at $|\Omega|\approx2|\alpha|$. As for the case of large $\beta$ ($\beta=1$) shown in Fig.~\ref{fig:Disorder-results-s-wave-4}(c), $\mbox{Re}\,\sigma$ resembles that in the normal phase (cf. with Fig.~\ref{fig:Disorder-results-s-wave-alpha=0}(a)). Indeed, such a result can be inferred from the spectral function and the DOS (see Fig.~\ref{fig:Disorder-Atr-s-wave-4}(c) and \ref{fig:Disorder-DOS-s-wave-4-2}(a)).
In the case of the intermediate values of $\beta$, the results also demonstrate the peaks, which are shifted toward smaller frequencies.

\begin{figure*}[!ht]
\begin{center}
\includegraphics[width=0.32\textwidth]{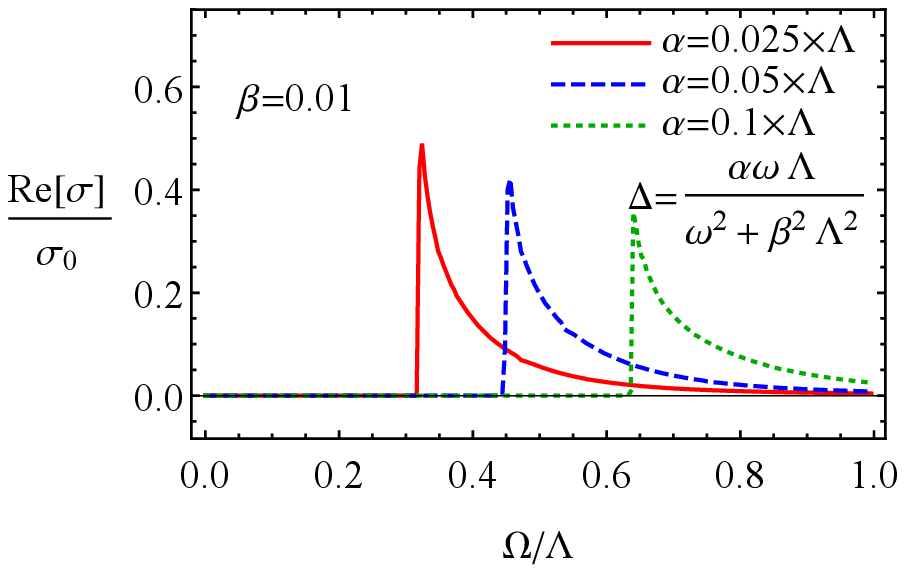}\hfill
\includegraphics[width=0.32\textwidth]{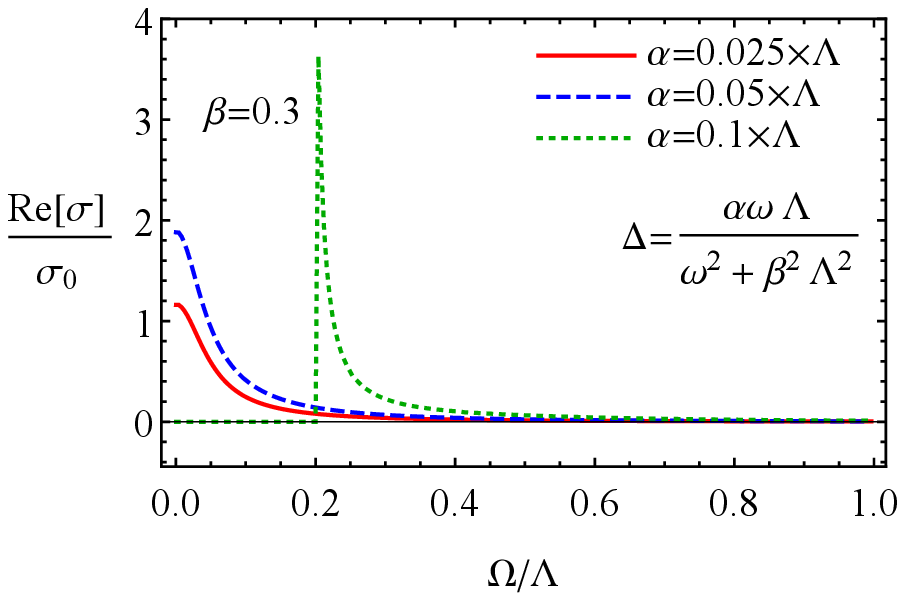}\hfill
\includegraphics[width=0.32\textwidth]{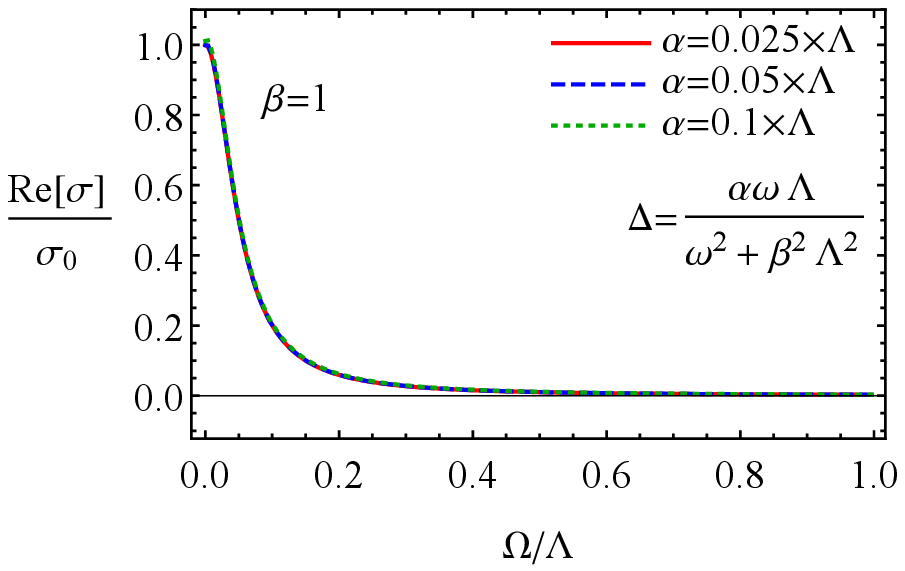}
\end{center}
\hspace{0.01\textwidth}{\small (a)}\hspace{0.3\textwidth}{\small (b)}\hspace{0.3\textwidth}{\small (c)}\\[0pt]
\caption{The real part of the optical conductivity for $\Delta= \alpha \omega /\left(\omega^2 +\beta^2 \Lambda^2\right)$ at a few values of $\alpha$. We set $\beta=0.01$ (panel (a)) and $\beta=0.3$ (panel (b)), and $\beta=1$ (panel (c)). In addition, $\tau=10/\Lambda$ and $T=0$ in all panels.
}
\label{fig:Disorder-results-s-wave-2}
\end{figure*}

The results for the other gap ansatz, i.e., $\Delta=\alpha \omega \Lambda/\left(\omega^2 +\beta^2 \Lambda^2\right)$, are shown in Fig.~\ref{fig:Disorder-results-s-wave-2} for a few values of $\beta$. The behavior of the gap at both small and large of the parameter $\beta$ qualitatively resembles that for $\Delta=\alpha \omega/\sqrt{\omega^2 +\beta^2 \Lambda^2}$, i.e., the peaks are absent and the dependence on the gap strength is weak. On the other hand, the position of the peaks are no longer given by $|\Omega|=2|\alpha|$ even at small $\beta$.
In addition, they generically have a smaller amplitude. In the case of intermediate values of $\beta$, the peak is present only for sufficiently strong gap and gradually shifts toward $\Omega=0$ with the decrease of $\alpha$.

It is worth noting that both realistic gap ansatzes have qualitatively the same optical conductivity for large values of $\beta$, which is weakly sensitive to the amplitude of the gap.
Therefore, while the OF pairing might be still present, it has almost no manifestations in the optical response and spectroscopic signals. This finding might explain why it is hard to experimentally find the signatures of the Berezinskii pairing.
As for the case of small values of $\beta$, the position and the amplitude of the observed attenuation peaks (i.e., the peaks in $\mbox{Re}\,\sigma$) strongly depend on the frequency profile of the gap. Moreover, they are generically higher and have a sharper form than their counterparts for the BCS pairing.
Thus, unlike the spectral function and the electron DOS, which fail to distinguish $\Delta=\alpha$ and $\Delta=\alpha \sign{\omega}$, the optical absorbtion quantified by $\mbox{Re}\,\sigma$ provides an alternative route to separate the Berezinskii and even-frequency pairings.

\subsection{Imaginary part of the optical conductivity}
\label{sec:Disorder-results-Im}

For completeness, we also discuss the imaginary part of the optical conductivity. To simplify the presentation, we consider the case of clean superconductors with the s-wave pairing, where $\mbox{Im}\,\sigma_{ij}$ is given in Eq.~(\ref{Disorder-sigma-Im-clean}). [This has an additional benefit because allows one to separate the effects of the disorder and OF pairing.]
Further, it is convenient to present the results normalized to the diamagnetic term in the optical conductivity (\ref{Disorder-sigma-Im-diam}), i.e., $\sigma^{\rm diam}\equiv \mbox{Im}\,\sigma_{ii}^{\rm diam}(\Omega)$.

\begin{figure*}[!ht]
\begin{center}
\includegraphics[width=0.45\textwidth]{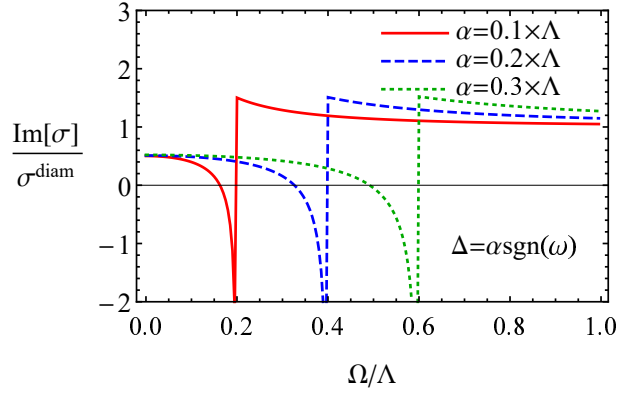}
\end{center}
\caption{The imaginary part of the optical conductivity $\mbox{Im}\,\sigma$ for $\Delta=\alpha \sign{\omega}$ at a few values of $\alpha$. We set $T=0$.}
\label{fig:Disorder-results-s-wave-Im-alpha=const-v1}
\end{figure*}

In the case of the simple BCS gap $\Delta=\alpha$, the imaginary part of the optical conductivity is trivial and is given by primarily diamagnetic term $\sigma^{\rm diam}$. The results for a simple OF gap $\Delta=\alpha \sign{\omega}$ are shown in Fig.~\ref{fig:Disorder-results-s-wave-Im-alpha=const-v1}. One can clearly see that there are cusp-like features where the imaginary part of the optical conductivity can change the sign. As expected from the Kramers-–Kronig relations, the cusps in the imaginary part of the optical conductivity reflect the sharp onsets in its real part, which occur at $|\alpha|=2|\Omega|$ (cf. Figs.~\ref{fig:Disorder-results-s-wave-alpha=0}(c) and  \ref{fig:Disorder-results-s-wave-Im-alpha=const-v1}).

\begin{figure*}[!ht]
\begin{center}
\includegraphics[width=0.32\textwidth]{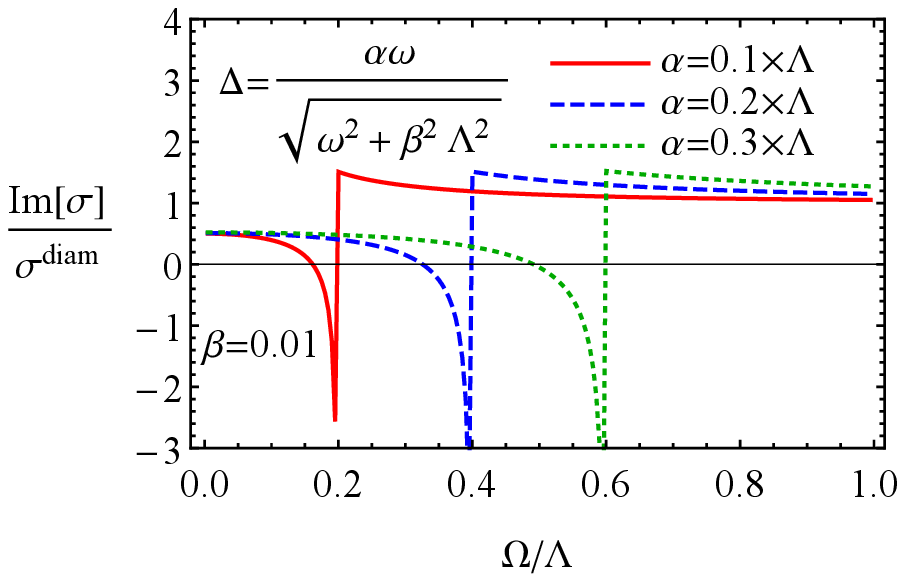}\hfill
\includegraphics[width=0.32\textwidth]{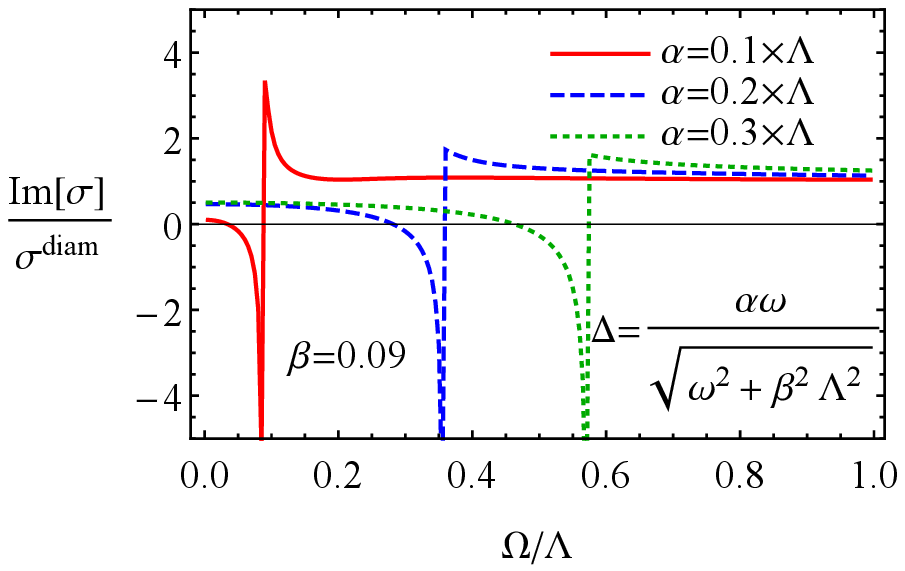}\hfill
\includegraphics[width=0.32\textwidth]{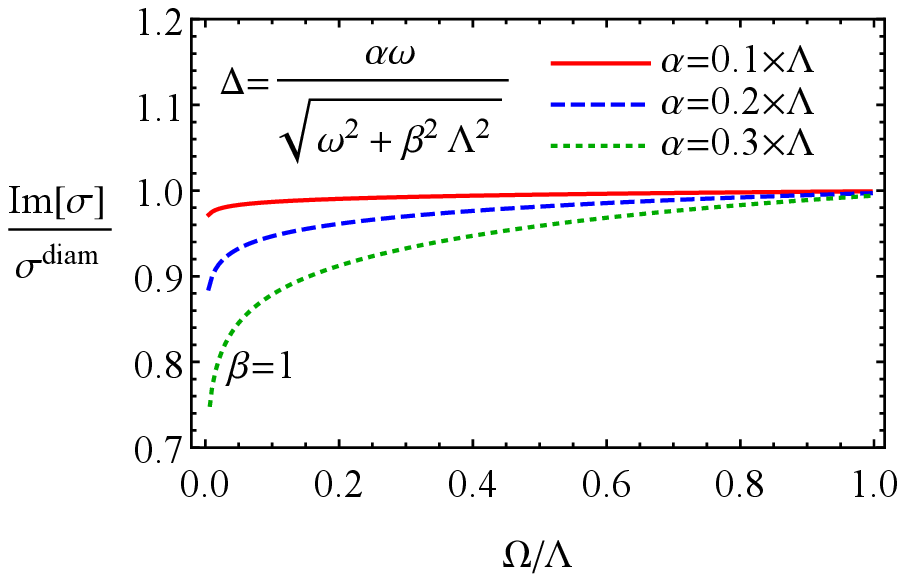}
\end{center}
\hspace{0.01\textwidth}{\small (a)}\hspace{0.3\textwidth}{\small (b)}\hspace{0.3\textwidth}{\small (c)}\\[0pt]
\caption{The imaginary part of the optical conductivity $\mbox{Im}\,\sigma$  for $\Delta= \alpha \omega/\sqrt{\omega^2 +\beta^2 \Lambda^2}$ at a few values of $\alpha$. We set $\beta=0.01$ (panel (a)), $\beta=0.09$ (panel (b)), and $\beta=1$ (panel (c)). In addition, $T=0$ in all panels.
}
\label{fig:Disorder-results-s-wave-Im-4-v1}
\end{figure*}

\begin{figure*}[!ht]
\begin{center}
\includegraphics[width=0.32\textwidth]{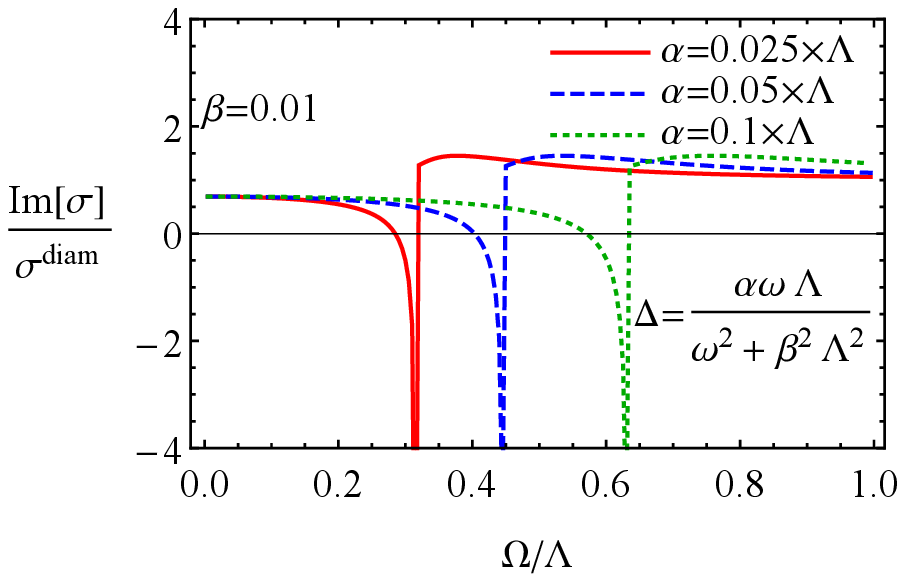}\hfill
\includegraphics[width=0.32\textwidth]{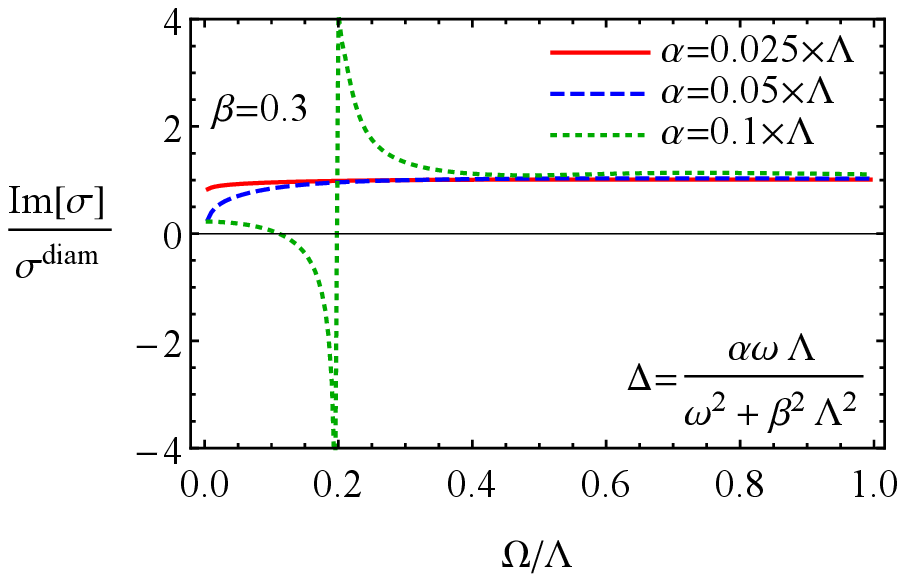}\hfill
\includegraphics[width=0.32\textwidth]{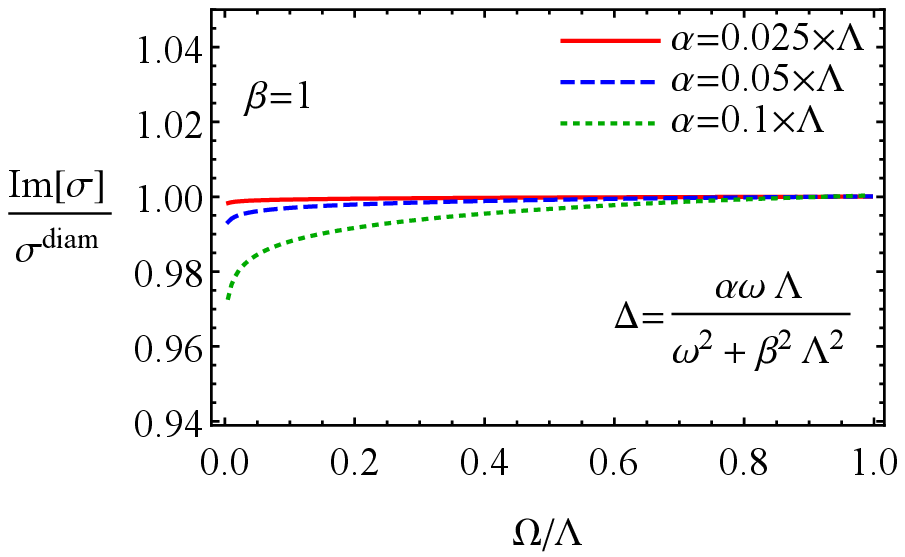}
\end{center}
\hspace{0.01\textwidth}{\small (a)}\hspace{0.3\textwidth}{\small (b)}\hspace{0.3\textwidth}{\small (c)}\\[0pt]
\caption{The imaginary part of the optical conductivity $\mbox{Im}\,\sigma$  for $\Delta= \alpha \omega \Lambda/\left(\omega^2 +\beta^2 \Lambda^2\right)$ at a few values of $\alpha$. We set $\beta=0.01$ (panel (a)), $\beta=0.3$ (panel (b)), and $\beta=1$ (panel (c)). In addition, $T=0$ in all panels.
}
\label{fig:Disorder-results-s-wave-Im-2-v1}
\end{figure*}

The imaginary part of the optical conductivity for $\Delta= \alpha \omega/\sqrt{\omega^2 +\beta^2 \Lambda^2}$ and $\Delta= \alpha \omega \Lambda/\left(\omega^2 +\beta^2 \Lambda^2\right)$ is shown in Figs.~\ref{fig:Disorder-results-s-wave-Im-4-v1} and \ref{fig:Disorder-results-s-wave-Im-2-v1}, respectively. As in the case of a simple gap $\Delta=\alpha \sign{\omega}$, the cusps in $\mbox{Im}\,\sigma$ correspond to the sharp onsets in the real part of the optical conductivity (see also Figs.~\ref{fig:Disorder-results-s-wave-4} and \ref{fig:Disorder-results-s-wave-2}). The peaks in the real part of the optical conductivity correspond to the relatively weak drops in $\mbox{Im}\,\sigma$ at small $\Omega$. The latter are clearly evident from Figs.~\ref{fig:Disorder-results-s-wave-Im-4-v1}(c), \ref{fig:Disorder-results-s-wave-Im-2-v1}(b), and \ref{fig:Disorder-results-s-wave-Im-2-v1}(c).

A common feature for all OF gaps is that the imaginary part of the optical conductivity can become negative for the frequencies preceding the appearance of the peaks in $\mbox{Re}\,\sigma$. For example, this is the case for $\Delta=\alpha\,\sign{\omega}$ at $|\Omega| \lesssim 2|\alpha|$, which is clearly evident from Fig.~\ref{fig:Disorder-results-s-wave-Im-alpha=const-v1}. The width of the corresponding region depends on the magnitude of the gap strength $\alpha$. In order to understand the physical meaning of this result, let us consider the dielectric permittivity $\varepsilon(\Omega)$. At small frequencies it reads as (see, e.g., Ref.~\cite{Mahan:book,Landau:t8})
\begin{equation}
\label{sec:Disorder-results-epsilon}
\varepsilon(\Omega)=\varepsilon_{\infty} + \frac{4\pi i \sigma(\Omega)}{\Omega} = \varepsilon_0 -\frac{4\pi \mbox{Im}\left[\sigma(\Omega)\right]}{\Omega} +i\frac{4\pi  \mbox{Re}\left[\sigma(\Omega)\right]}{\Omega},
\end{equation}
where $\varepsilon_{\infty}=\lim_{\Omega\to\infty}\varepsilon(\Omega)$.
Obviously, $\mbox{Re}\left[\sigma(\Omega)\right]$ should be positive to describe the dissipation of electromagnetic waves inside the media. On the other hand, there is no physical restriction on the sign of the imaginary part (see, e.g., Ref.~\cite{Landau:t8}). In particular, while $\mbox{Im}\left[\sigma(\Omega)\right]>0$ corresponds to the reflection of an electromagnetic wave with sufficiently small frequency, the negative imaginary part $\mbox{Im}\left[\sigma(\Omega)\right]<0$ increases the effective refractive index and allows for the propagation of light. To the best of our knowledge, such a feature does not appear in conventional even-frequency superconductors. Therefore, the observation of these \emph{transparency windows} on the onset of the attenuation peaks (i.e., the peaks in the real part of the optical conductivity) might be a promising experimental signature of the Berezinskii pairing. Last but not least, we note that the zeros of dielectric function (\ref{sec:Disorder-results-epsilon}) corresponds to plasmon excitations. While they could in principle exist at small frequencies, the plasmons should be strongly damped at large frequencies, e.g., $|\Omega| \gtrsim 2|\alpha|$, where the attenuation peaks lead to $\mbox{Im}\,\varepsilon(\Omega)\neq0$.

\section{Summary}
\label{sec:Summary}

In this study, we investigated the spectroscopic signatures of the Berezinskii (or, equivalently odd-frequency) pairing in a simple model of an OF superconductor with a large Fermi surface and a parabolic energy spectrum.
The results are obtained for a few ansatzes of the OF superconducting gaps. In addition, a simple non-magnetic disorder is taken into account. While the general expressions for the spectral function, the DOS, and the optical conductivity are valid for an arbitrary dependence of the gap on momentum, we focus on the case of the s-wave pairing.

The analysis of the spectral function and the electron density of states showed that, unlike naive expectations, the OF pairing can be also manifested as the spectral gap. This is explained by the fact that the poles of the Green function are found self-consistently and do not correspond to any quasiparticle states for small frequencies. The width of the spectral gap depends not only on the strength of the gap but also the frequency profile of the OF gaps.

In particular, the spectral gap closes for a relatively weak frequency dependence and the density of states could be enhanced at small frequencies followed by a local minimum.
In general, however, it might be challenging to distinguish the Berezinskii pairing from conventional BCS-like superconducting states. Depending on the frequency profile of the gaps, density of states can resemble even the normal state. Indeed, the coherence peaks observed for the OF gaps are qualitatively similar to those for even-frequency gaps and the observation of local minimum might be spoiled by various experimental uncertainties.

On the other hand, the manifestations of the OF gaps in the optical conductivity are rather powerful and unambiguous. In particular, the real part of the optical conductivity $\mbox{Re}\,\sigma$ might demonstrate an almost Drude-like behavior for a sufficiently weak dependence of an OF gap on frequency.
In addition, the conductivity in this regime depends weakly on the value of the gap. On the other hand, like in the case of the conventional BCS pairing, peaks at certain values of frequency appear in the optical response when an OF gap quickly changes with frequency. The magnitude of these peaks and their position are determined by the frequency profile of the gap. In addition, they are sharper compared to expected peaks for the conventional pairing. Therefore, we believe that the study of the optical absorption quantified by $\mbox{Re}\,\sigma$ might be a promising way in the search of the Berezinskii pairing.

The OF gaps have interesting manifestations in the imaginary part of the optical conductivity $\mbox{Im}\,\sigma$. Unlike the case of the BCS coupling, the sharp cusp-like features appear at the onsets of the peaks in $\mbox{Re}\,\sigma$. Moreover, it is found that the imaginary part might become negative in close proximity of cusps. Physically, this suggests that OF superconductors might allow for the optical transparency windows (i.e., regions of the parameter space where the reflection is negligible) for certain frequencies. This effect can be also used to effectively detect the elusive Berezinskii pairing phase in experiments. In the case of flat frequency profiles of the OF gaps, where there are no attenuation peaks, $\mbox{Im}\,\sigma$ demonstrates a weak reduction at small frequencies. The latter is in agreement with the enhancement of the real part of the conductivity.

Let us also discuss a few limitations of this study. While the model and the approximation of a large Fermi surface significantly simplify the calculations, a natural extension will be to study the case of Dirac and Weyl semimetals near the charge neutrality point. The nontrivial spin texture and the momentum-dependence of the gap are other ingredients that would need to be included. In addition, in order to study the effect of temperature on the optical conductivity, the dependence of the OF gap on temperature itself should be clarified. The latter, as well as the rigorous determination of the parameter $Z(\omega,\mathbf{p})$, requires the self-consistent treatment of the gap equations. These questions are, however, outside the scope of the current research and will be reported elsewhere.

\begin{acknowledgments}
We are grateful to E.~Langmann, J-M.~Tremblay and his group, and M.~Geilhufe for useful discussions. This work was supported by the VILLUM FONDEN via the Centre of Excellence for Dirac Materials (Grant No. 11744), the European Research Council under the European Unions Seventh Framework ERS-2018-SYG 810451 HERO and the Knut and Alice Wallenberg Foundation.
\end{acknowledgments}

\appendix

\section{Sum rule and $Z(\omega, \mathbf{p})$}
\label{sec:App-Sum-rules}

In this appendix, we discuss the sum rule in the case of the odd-frequency (OF) pairing. It is well known that the opening of the gap leads to the redistribution of the electron density of states (DOS) in usual superconductors. At the same time, the total spectral weight is conserved. Mathematically, this property is formulated in terms of the so-called sum rule. According to Refs.~\cite{Bruus-book,Altland:book,Mahan:book}, the electron part of the spectral function $\hat{A}(\omega,\mathbf{p})$ in the Nambu space integrated over all frequencies, i.e.,
\begin{eqnarray}
\label{App-OC-sum-rules-IA}
I_{\rm A} = \int_{-\infty}^{\infty}d\omega\, \mbox{tr}\left[\frac{1+\tau_z}{2}\hat{A}(\omega,\mathbf{p})\right],
\end{eqnarray}
should satisfy
\begin{eqnarray}
\label{App-OC-sum-rules-AG}
I_{\rm A} = 1,
\end{eqnarray}
where the spectral function is defined in Eq.~(\ref{Disorder-SP-A-def}) and its explicit form is given in Eq.~(\ref{Disorder-SP-A}) in the main text.

\begin{figure*}[!ht]
\begin{center}
\includegraphics[width=0.45\textwidth]{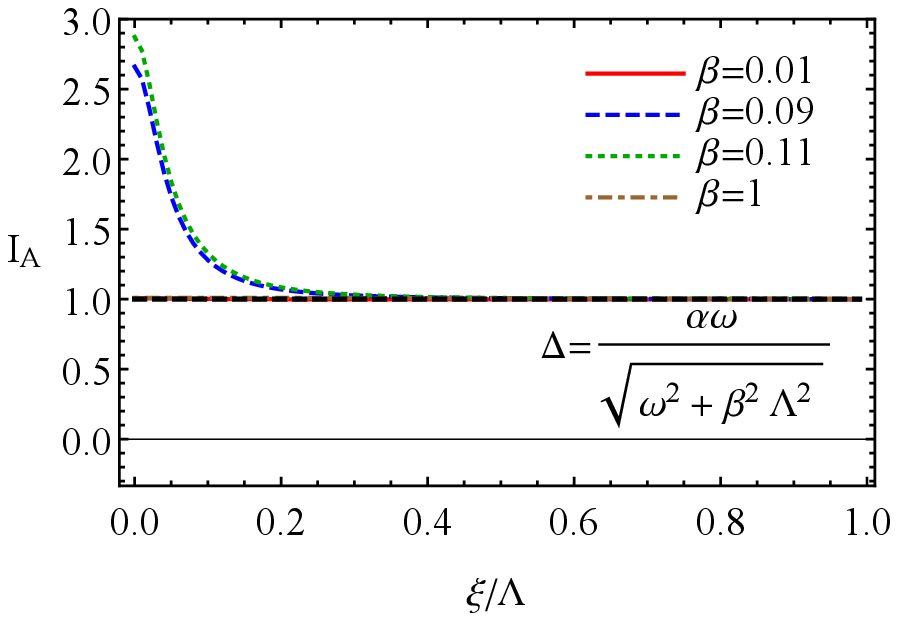}\hfill
\includegraphics[width=0.45\textwidth]{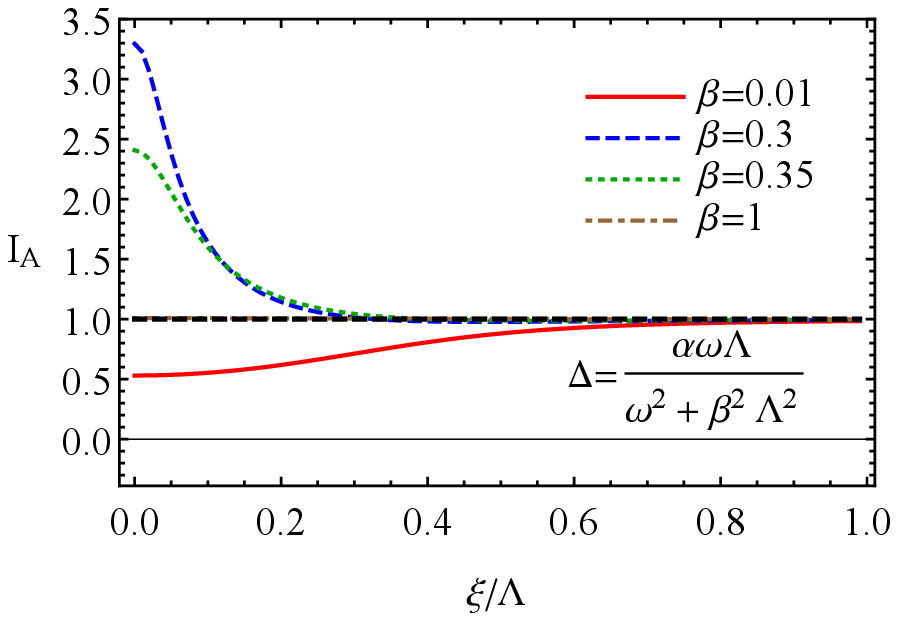}
\end{center}
\hspace{0.075\textwidth}{\small (a)}\hspace{0.525\textwidth}{\small (b)}\\[0pt]
\caption{The integral over the frequency from the spectral function $I_{\rm A}$ defined in Eq.~(\ref{App-OC-sum-rules-IA}) for the OF gaps $\Delta=\alpha \omega/\sqrt{\omega^2 +\beta^2 \Lambda^2}$ and $\Delta=\alpha \omega \Lambda/\left(\omega^2 +\beta^2 \Lambda^2\right)$ at a few values of $\beta$. The sum rule (\ref{App-OC-sum-rules-AG}) reads as $I_{\rm A}=1$ and is denoted by the black solid line. We set $Z(\omega,\mathbf{p})=1$, $\tau=10/\Lambda$, $T=0$, and $\alpha=0.1\Lambda$ in both panels.}
\label{fig:Disorder-DOS-s-wave-sum-rule}
\end{figure*}

In order to show the importance of the additional term $Z(\omega,\mathbf{p})$ in the Green function (\ref{OC-SF-G-BdG-def}), let us check whether the sum rule (\ref{App-OC-sum-rules-AG}) holds if this term is ignored. It is straightforward to verify that the sum rule (\ref{App-OC-sum-rules-AG}) is not broken by neither conventional BCS $\Delta=\alpha$ nor simplest odd-frequency gap $\Delta=\alpha\sign{\omega}$. However, this is generically not the case for other gap ansatzes. The corresponding results for the OF gap $\Delta=\alpha \omega/\sqrt{\omega^2 +\beta^2 \Lambda^2}$ at a few values of the control parameter $\beta$ are shown in Fig.~\ref{fig:Disorder-DOS-s-wave-sum-rule}(a). As one can see, the deviation from unity is very small for both large (e.g., $\beta\approx 1$) and small values of $\beta$, where the electron DOS behaves similarly to the normal and usual BCS phases, respectively. For intermediate $\beta$, the sum rule is, however, noticeably broken. As for the OF gap $\Delta=\alpha \omega \Lambda/\left(\omega^2 +\beta^2 \Lambda^2\right)$, the sum rule in Fig.~\ref{fig:Disorder-DOS-s-wave-sum-rule}(b) is satisfied only at large $\beta$ (e.g., $\beta\approx 1$) and is broken even at $\beta\to0$.

The breakdown of the sum rule (\ref{App-OC-sum-rules-AG}) is particularly strong when the gap in the spectral function and the DOS disappear (see the results in Sec.~\ref{sec:DOS} in the main text) reaching the maximal value exactly at the merging point (e.g., $\beta=\alpha$ for $\Delta=\alpha \omega/\sqrt{\omega^2 +\beta^2 \Lambda^2}$). With the increase of $\beta$, however, the sum rule is quickly restored for both OF gaps $\Delta=\alpha \omega/\sqrt{\omega^2 +\beta^2 \Lambda^2}$ and $\Delta=\alpha \omega \Lambda/\left(\omega^2 +\beta^2 \Lambda^2\right)$. Therefore, since the breakdown of the sum rule usually signifies an inconsistency of model, the additional term $Z(\omega,\mathbf{p})$ should be taken into account in the calculation of the spectroscopic properties of the OF superconductors. As we argued in the main text, this term might indeed appear in the self-consistent Eliashberg approach and is the renormalization of the wave-function.

\begin{figure*}[!ht]
\begin{center}
\includegraphics[width=0.45\textwidth]{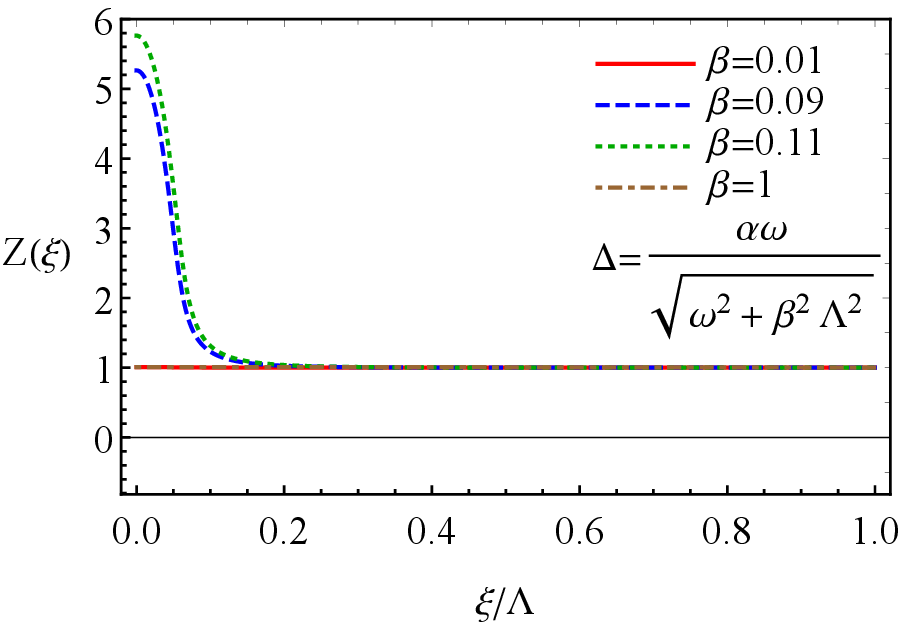}\hfill
\includegraphics[width=0.45\textwidth]{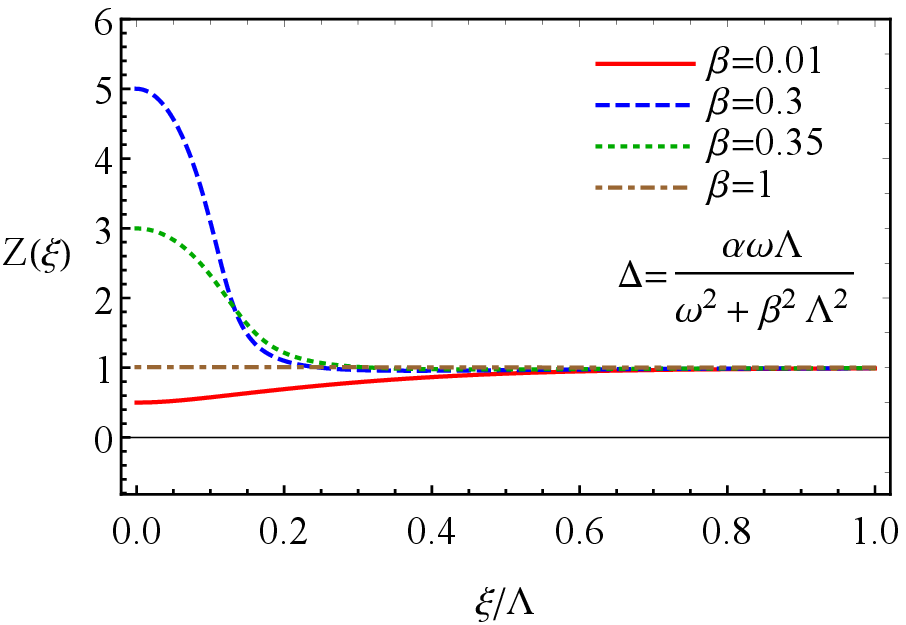}
\end{center}
\hspace{0.075\textwidth}{\small (a)}\hspace{0.525\textwidth}{\small (b)}\\[0pt]
\caption{The coefficient $Z(\omega, \mathbf{p}) \approx Z(\xi)$ as a function of the energy $\xi$ for $\Delta=\alpha \omega/\sqrt{\omega^2 +\beta^2 \Lambda^2}$ (panel (a)) and $\Delta=\alpha \omega \Lambda/\left(\omega^2 +\beta^2 \Lambda^2\right)$ (panel (b)). We set $\tau\to \infty$, $\alpha=0.1\Lambda$, and $T=0$ in both panels.
}
\label{fig:Z-s-wave-4-2}
\end{figure*}

In passing, we present the coefficient $Z(\omega,\mathbf{p})\approx Z(\xi)$ in the clean case. As is discussed in Sec.~\ref{sec:DOS-A} in the main text, it is obtained by considering the sum rule (\ref{App-OC-sum-rules-AG}) as an integral equation for $Z(\xi)$. The corresponding results are shown in Fig.~\ref{fig:Z-s-wave-4-2}. The results for the dirty case are presented in Fig.~\ref{fig:Z-s-wave-4-2-eta} in the main text. In order to mitigate the breakdown of the sum rule at small $\Omega$, the coefficient $Z(\xi)$ clearly correlates to the sum rule breakdown in Fig.~\ref{fig:Disorder-DOS-s-wave-sum-rule} reaching the maximal value at small frequencies. It is worth noting that effects of disorder are quantitative rather than qualitative.

\end{document}